\documentclass[fleqn,usenatbib]{mnras}

\usepackage{newtxtext,newtxmath}

\usepackage[T1]{fontenc}
\usepackage{ae,aecompl}

\usepackage{graphicx}	
\usepackage{amsmath}	
\usepackage{amssymb}	

\usepackage{xspace}
\usepackage[normalem]{ulem}

\newcommand{\CI}{[CI]\xspace}
\newcommand{\CII}{[CII]\xspace}
\newcommand{\NII}{[NII]\xspace}

\def\h2{\ifmmode {\mbox H$_2$}\else H$_2$\fi\xspace}

\newcommand*{\img}[1]{%
\raisebox{-.4\baselineskip}{%
        \includegraphics[scale = 0.05]{#1}%
    }%
}

\title[The art of modeling CO, \CI, and \CII \img{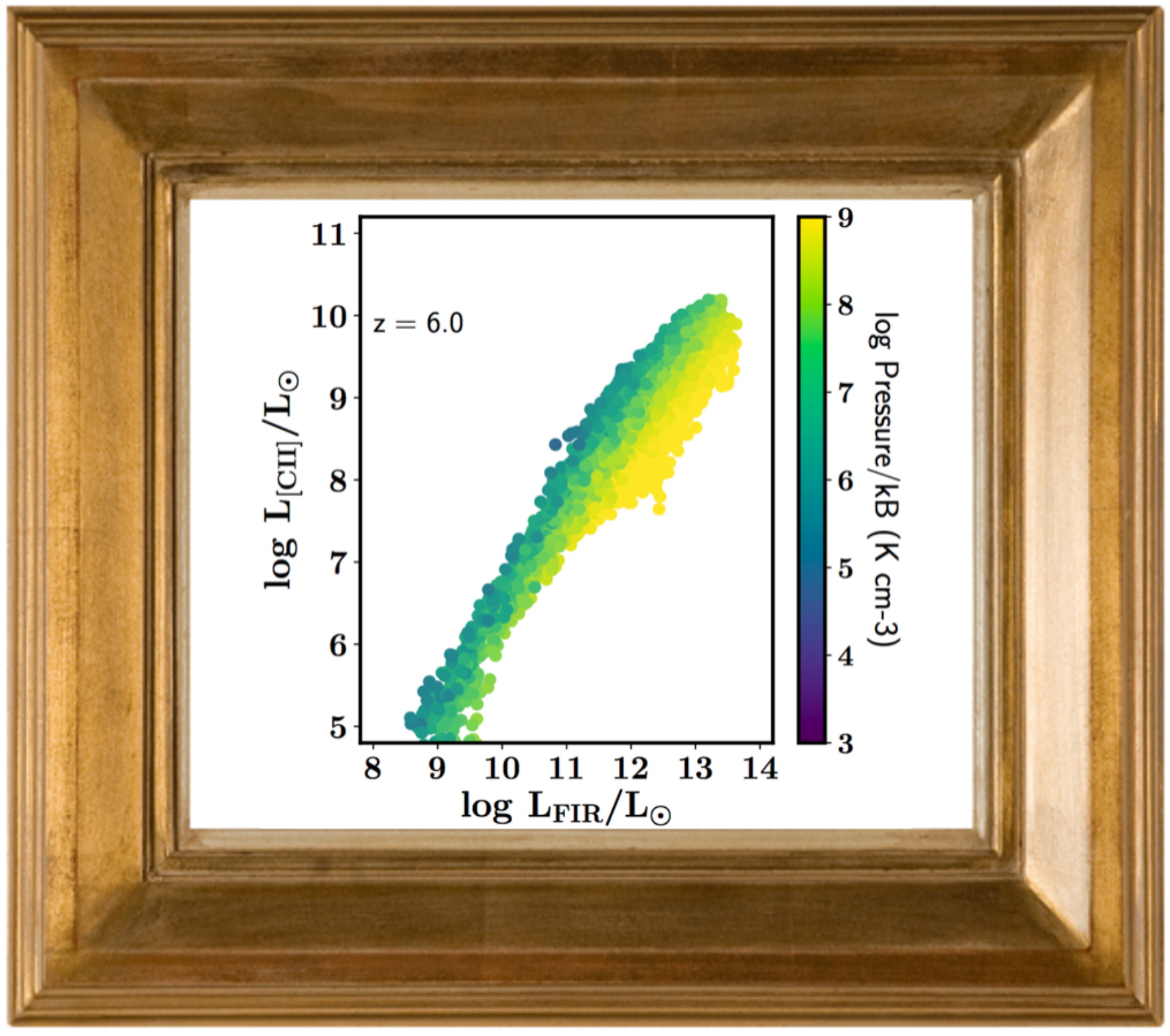}]{The art of modeling CO, \CI, and \CII in cosmological galaxy formation models}

\author[G. Popping et al.]{Gerg\"o Popping,$^{1}$\thanks{MPIA fellow, e-mail: popping@mpia.de} 
Desika Narayanan,$^{2,3,4}$ Rachel S. Somerville,$^{5,6}$ Andreas
L. Faisst,$^{7}$ \newauthor and Mark
R. Krumholz$^{8,9}$  \\
$^{1}$Max-Planck-Institut f\"ur Astronomie, K\"onigstuhl 17, D-69117 Heidelberg, Germany\\
$^{2}$Department of Astronomy, University of Florida, 211 Bryant Space Science Center, Gainesville, FL 32611, USA\\
$^{3}$University of Florida Informatics Institute, 432 Newell Drive, CISE Bldg E251, Gainesville, FL 32611\\
$^{4}$Cosmic Dawn Center (DAWN), Niels Bohr Institute, University of Copenhagen, Juliane Maries vej 30, DK-2100 Copenhagen, Denmark\\
$^{5}$Center for Computational Astrophysics, Flatiron Institute, 162 5th Ave, New York, NY 10010, USA\\
$^{6}$Department of Physics and Astronomy, Rutgers University, 136 Frelinghuysen Road, Piscataway, NJ 08854, USA\\
$^{7}$IPAC, Mail Code 314-6, California Institute of Technology, 1200 East California Boulevard, Pasadena, CA 91125, USA\\
$^{8}$Research School of Astronomy and Astrophysics, Australian National University, Canberra 2611, ACT, Australia\\
$^{9}$Centre of Excellence for Astronomy in Three Dimensions (ASTRO-3D), Australia\\
}

\date{Accepted XXX. Received YYY; in original form ZZZ}

\pubyear{2018}
\hypersetup{draft}
\begin{document}
\label{firstpage}
\pagerange{\pageref{firstpage}--\pageref{lastpage}}
\maketitle

\begin{abstract}
The advent of new sub-millimeter observational facilities has
stimulated the desire to model the sub-mm line emission of galaxies
within cosmological galaxy formation models. This is typically done by applying sub-resolution recipes to describe the
properties of the unresolved interstellar medium.  While there is freedom in how one implements sub-resolution recipes, the impact of various choices has yet to be
  systematically explored.  We combine a semi-analytic model of galaxy formation with
chemical equilibrium networks and numerical radiative transfer models
and explore how different choices for the sub-resolution modeling affect the
predicted CO, \CI, and \CII emission of galaxies. A key component for a successful model includes a molecular cloud
mass--size relation and scaling for the ultraviolet and cosmic ray
radiation field that depend on local ISM properties. Our most
successful model adopts a Plummer radial density profile for gas within molecular clouds. Different
assumptions for the clumping of gas within molecular clouds and
changes in the molecular cloud mass distribution function
hardly affect the CO, \CI, and \CII luminosities of galaxies. At fixed
star-formation rate the \CII--SFR ratio of galaxies scales
inversely with the pressure acting on molecular clouds, increasing the
molecular clouds density and hence decreasing the importance of \CII
line cooling. We find that it is essential that a wide range
of sub-mm emission lines arising in vastly different phases of the ISM
are used as model constraints in order to limit the freedom in
sub-grid choices.
\end{abstract}




\section{Introduction}
Sub-millimeter astronomy has grown significantly over the last decade
with the advent of new and improved instruments such as the Atacama
Large (sub-)Millimeter Array (ALMA), the NOrthern Extended Millimeter
Array (NOEMA), and the Large Millimeter Telescope (LMT). This field is
expected to grow even further once new instruments such as CCAT-prime
and the currently discussed new instruments such as the next-generation Very Large Array (ngVLA) and the Atacama Large-Aperture Submm/mm Telescope (AtLAST) come on-line. The quick rise in sub-mm collecting area and sensitivity has enabled the efficient collection of sub-mm emission line information for large numbers of galaxies over cosmic time \citep[see reviews by][]{carilli13a,casey14a}. 

At the same time, the available and expected observations from the newest generation of sub-mm facilities present a new and stringent challenge to theoretical models of galaxy formation. In particular, the rapidly growing number of CO \citep[e.g.,][]{Daddi2010,Aravena2014,Tacconi2013,Walter2016,Decarli2016a,Papovich2016,Tacconi2017}, \CI \citep[e.g.,][]{Bothwell2017,Popping2017csfg} and \CII \citep[e.g.,][]{Brisbin2015,Capak2015,Schaerer2015,Knudsen2016,Inoue2016} 
detections at $z>0$ place a strong constraint on the ISM phase structure within galaxy formation models \citep[see for a recent review][and compilations presented in for example \citet{Tacconi2017} and \citet{Olsen2017}]{carilli13a}.  As a result, there has been significant interest within the galaxy formation community in modeling physics of these line emission processes within galaxy formation simulations.

The main challenges when predicting the sub-mm line emission from
galaxy formation models is the large dynamic range of physical scales
that have to be addressed. A successful model simultaneously needs to
address galaxy baryonic physics acting on Mpc scales (or even larger
cosmological scales), kpc and pc scales for the distribution of matter
within galaxies and the physics acting upon this matter, and atomic
physics on sub-pc scales within molecular clouds. Combining these
scales within one model is not computationally feasible, which has
made theorists resort to ``sub-resolution approaches'' (also called
"sub-grid"). Developing these sub-grid approaches is not always
straightforward and is usually based on either high-resolution idealised
simulations or observations. In this paper we do not discuss the
sub-grid recipes invoked to describe physical processes acting on the
baryons in galaxies \citep[e.g., star-formation, stellar and active galactic
nuclei feedback,][]{Somerville2014}. Instead we focus on the key
sub-grid choices that are relevant in the context of modeling sub-mm
line emission from galaxies in post-processing. This includes assumptions for the distribution and density profiles of molecular clouds, the radiation field, and the treatment of ionized gas.

Over the last decade multiple groups have focused on the modeling of
sub-mm emission lines such as CO, \CI, and \CII from galaxies, either
based on semi-analytic galaxy formation models
\citep{Lagos2012,Popping2014CO,Popping2016, Lagache2017}, hydrodynamic
models \citep{nagamine06a,Narayanan2008,narayanan11b,Narayanan2012,Narayanan2014,Olsen2015CO,Olsen2015CII,Vallini2015,Olsen2017,Katz2017,Pallottini2017,Vallini2018},
or analytic models \citep{Narayanan2017,Munoz2013,Munoz2016}. All these
groups used a (cosmological) galaxy formation model as a starting
point and combined this with machinery to model the sub-mm line
emission of galaxies in post-processing. This machinery usually
includes the coupling to a spectral synthesis code such as
\texttt{CLOUDY} \citep{CLOUDY} or a photo-dissociation region (PDR)
code such as \texttt{DESPOTIC} \citep{Krumholz2013,Krumholz2014}. An
additional essential part of this machinery is the previously
discussed sub-grid choices for the structure of the ISM. Sub-resolution choices ranging from
  imposed floor or fixed densities to varying density profiles
  (e.g. logotropic, Plummer, power-law and constant) to 
  varying molecular cloud mass functions to diverse clumping
  factors have all been assumed within the literature
\citep[e.g.][]{Lagos2012,Narayanan2012,Popping2014,Popping2016,Olsen2017,Vallini2018}.

Despite the wide range in assumptions that have been made for the
sub-grid modeling, all these groups have successfully reproduced the
sub-mm line emission of galaxies compared to observational
constraints. This demonstrates that there is still a lot of freedom in
the choices one can make for the sub-grid physics. These efforts have typically only focused on the emission from
one molecule or atom \citep[e.g., only CO or only \CII emission, although see][]{Olsen2017,Pallottini2017}. That
  said the emission from different atomic or molecular species can
  arise from drastically different ISM physical conditions.  For
  example, $^{12}$CO (hereafter, CO) typically is associated with
  molecular H$_2$ gas, while atomic [CI] can come from both molecular
  and neutral gas.  Even more extreme is [CII] emission (emitted by
  singly ionized carbon, C$^+$), which can
  reside cospatially with molecular, neutral or ionized hydrogen. A
model that successfully reproduces the \CII emission of galaxies
therefore does not necessarily reproduce the emission from a molecular
ISM tracer such as CO or HCN as well. Successfully reproducing the
emission from multiple atoms and molecules simultaneously is therefore
more challenging and has the potential to narrow down the freedom in
designing the sub-grid approaches.

 A systematic study of the typical
  choices made in sub-resolution modeling and their effect on the observed sub-mm line
  properties is thus important. In this paper we
explore how different sub-grid choices to represent the ISM in
galaxies affect the resulting CO, \CI and \CII emission of
galaxies, while keeping the underlying galaxy formation model fixed \citep[other works have also assessed the impact of some of their sub-resolution prescriptions, e.g.,][]{Olsen2017,Vallini2018}. As a starting point we use a semi-analytic model of galaxy
formation. We explore various sub-grid approaches to describe the
distribution of diffuse and dense gas within the ISM of galaxies,
especially focusing on the mass distribution function of molecular
clouds, the density distribution profile within molecular clouds,
clumping within molecular clouds, the ultraviolet and cosmic ray field
impinging on molecular clouds, and the treatment of ionized gas. We
combine chemical equilibrium networks and numerical radiative transfer
models with sub-grid models to develop a picture of how the emission
of CO, \CI, and \CII changes within galaxies. We aim to explore if the
freedom in sub-grid assumptions can be limited when using a
combination of multiple sub-mm emission lines as model constraints and
try converge to a fiducial model that best reproduces the CO, \CI, and
\CII emission of galaxies simultaneously. We do not aim to derive the characteristics (e.g., density profile) of giant molecular clouds in galaxies. We rather aim to find an operational prescription for the sub-mm emission of galaxies. Our conclusions about which model agrees best with observations are of course sensitive to the
predicted "underlying" properties from our particular SAM. While these conclusions may be fairly sensitive to the specifics of the galaxy formation model, the conclusions regarding how the details of the sub-grid modeling impacts the sub-mm line observables are robust.

This paper is structured as followed. In Section \ref{sec:model} we
describe the model followed by a brief description of how different
sub-grid choices affect the carbon chemistry in molecular clouds
(Section \ref{sec:carbon_chemistry}). In Section \ref{sec:results} we describe the main results, while we discuss these in Section \ref{sec:discussion}. We summarise our main results and conclusions in Section \ref{sec:conclusions}.  Throughout this paper we adopt a flat $\Lambda$CDM
cosmology with $\Omega_0=0.28$, $\Omega_\Lambda = 0.72$,
$h=H_0/(100\,\rm{km}\,\rm{s}^{-1}\,\rm{Mpc}^{-1}) = 0.7$,
$\sigma_8=0.812$, and a cosmic baryon fraction of $f_b=0.1658$
\citep{Komatsu2009} and a Charbier \citep{Chabrier2003} initial mass function.

\section{Models}
\label{sec:model}
\subsection{Galaxy formation model}
\label{sec:sam}
We use the 'Santa Cruz' semi-analytic galaxy formation model \citep{Somerville1999,Somerville2001} as the underlying galaxy formation model in this paper. Significant updates
to this model are described in \citet{Somerville2008},
\citet{Somerville2012}, \citet{Porter2014}, \citet[from here on PST14]{Popping2014}, and \citet[from here on SPT15]{Somerville2015}. The model tracks the hierarchical clustering
of dark matter haloes, shock heating and radiative cooling of gas, SN feedback, SF, AGN
feedback (by quasars and radio jets), metal enrichment of the
interstellar and intracluster medium, mergers of galaxies, starbursts,
the evolution of stellar populations, the growht of stellar and gaseous disks, and dust obscuration, as well as the abundance of ionized, atomic and molecular hydrogen and a molecular hydrogen-based star-formation recipe. In this section we briefly summarise recipes that are important components of the model with regards to the modeling of sub-mm emission lines (recipes to track the ionized, atomic, and molecular hydrogen abundance and the molecule-based SF-recipe). We point the reader to \citet{Somerville2008}, \citet{Somerville2012}, PST14, and SPT15 for a more detailed description of the model. 

The cold gas component of a galaxy within the SAM consists of an ionized, atomic and
molecular component (as outlined in PST14 and SPT15). The ionized component may be ionized either by an external background or by the radiation field from stars within the
galaxy (a fixed fraction $f_{\rm ion, int}$).  The
external background field ionizes a slab of gas on each side of the
disc. Assuming that all the
gas with a surface density below some critical value $\Sigma_{\rm
  HII}$ is ionized by the external background, we  write \citep{Gnedin2012}
\begin{equation}
 f_{\rm ion} = \frac{\Sigma_{\rm HII}}{\Sigma_0}
\left[1 + \ln \left(\frac{\Sigma_0}{\Sigma_{\rm HII}} \right) + 0.5
  \left(\ln \left(\frac{\Sigma_0}{\Sigma_{\rm HII}}\right) \right)^2
  \right]. 
\end{equation}
Supported by the results of \citet{Gnedin2012} we assume throughout this paper $f_{\rm ion, int} = 0.2$ (as in the
Milky Way) and $\Sigma_{\rm HII} = 0.4 \, M_\odot \rm{pc}^{-2}$.

The \h2 fraction of the cold gas is computed based on the work by \citet{Gnedin2011}. The authors performed high-resolution `zoom-in' cosmological simulations including gravity,
hydrodynamics, non-equilibrium chemistry, and simplified 3D
on-the-fly radiative transfer.  They find that the \h2 fraction of the cold gas can be described by a simple fitting formula as a function of  the
dust-to-gas ratio relative to solar, $D_{\rm MW}$, the ionizing background radiation
field, $U_{\rm MW}$, and the surface density of the cold gas,
 $\Sigma_{\rm HI + H2}$. 
The described the molecular hydrogen fraction as
\begin{equation}
\label{eq:H2_GK}
 f_{H_2} = \left[1+\frac{\tilde{\Sigma}}{\Sigma_{HI+H_2}}\right]^{-2} 
\end{equation}
where
\begin{eqnarray*}
\tilde{\Sigma}  & = &  20\, {\rm M_\odot pc^{-2}} \frac{\Lambda^{4/7}}{D_{\rm MW}} 
\frac{1}{\sqrt{1+U_{\rm MW} D_{\rm MW}^2}}, \\
\Lambda & = & \ln(1+g D_{\rm MW}^{3/7}(U_{\rm MW}/15)^{4/7}),\\
g & = & \frac{1+\alpha s + s^2}{1+s},\\
s &  = & \frac{0.04}{D_*+D_{\rm MW}},\\
\alpha &  = & 5 \frac{U_{\rm MW}/2}{1+(U_{\rm MW}/2)^2},\\
D_* & = & 1.5 \times 10^{-3} \, \ln(1+(3U_{\rm MW})^{1.7}).
\end{eqnarray*}
In this work we assume that the dust-to-gas ratio is proportional to the
metallicity of the gas in solar units $D_{\rm MW} = Z_{\rm gas}/Z_\odot$. We
assume that the local ultraviolet (UV) background scales with the star-formation rate (SFR) relative to
the Milky Way value, $U_{\rm MW} = SFR/SFR_{\rm MW}$, where we choose
$SFR_{\rm MW} = 1.0\,\rm{M}_\odot\,\rm{yr}^{-1}$
\citep{Murray2010,Robitaille2010}. \citet{Popping2017dust} included the tracking of dust in the Santa Cruz galaxy formation model. In a future paper we will make our models self-consistent by instead
using the modeled dust abundance rather than gas-phase metallicity to estimate the molecular hydrogen fraction.

The star-formation recipe in the Santa Cruz SAM is based on an empirical relationship
between the surface density of molecular hydrogen and the surface
density of star-formation
\citep{Bigiel2008,Genzel2010,Bigiel2012}, accounting for an increased star-formation efficiency in environments with high molecular-hydrogen surface densities \citep[see PST14 and SPT15 for details]{Sharon2014,Hodge2014}. To following expression is used to model star formation 
\begin{equation}
\label{eqn:bigiel2}
\Sigma_{\rm SFR} = A_{\rm SF} \, \Sigma_{\rm H_2}/(10 M_\odot {\rm pc}^{-2}) \left(1+
\frac{\Sigma_{H_2}}{\Sigma_{\rm H_2, crit}}\right)^{N_{\rm SF}},
\end{equation}
where $\Sigma_{\rm H_2}$ is the surface density of molecular hydrogen
and with $A_{\rm
  SF}=5.98\times 10^{-3}\, M_\odot {\rm yr}^{-1} {\rm kpc}^{-2}$,
$\Sigma_{\rm H_2, crit} = 70 M_\odot$ pc$^{-2}$, and $N_{\rm
  SF}=1$. 

For this work, we construct the merging histories (or merger trees) of dark matter haloes based on the extended Press--Schechter formalism following the method described in \citet{SomervilleKolatt1999}, with improvements described in S08. 
We prefer EPS
merger trees in this work because they allow us to achieve high mass resolution,
useful to explore differences in the sub-grid approaches for low-mass
galaxies \citep[nearly identical results are obtained for our SAM when run on merger trees extracted from N-body simulations and on EPS merger trees;][]{Lu2014,Porter2014}. Haloes are resolved down to a minimum
progenitor mass $M_{\rm res}$ of $M_{\rm res} = 10^{10}\,\rm{M}_\odot$
for all root haloes, where $M_{\rm res}$ is the mass of the root halo
and represents the halo mass at the output redshift. A minimum resolution of $M_{\rm res} = 0.01 M_{\rm root}$ is imposed (see
Appendix A of \citet{Somerville2015} for more details on this minimum mass resolution). The simulations were run on a grid of haloes with root halo
masses ranging from $5\times 10^8$ to $5 \times 10^{14}\,\rm{M}_\odot$
at each redshift of interest, with 100 random realisations at
each halo mass. 
We have kept the galaxy formation parameters fixed to the
values presented in PST14 and SPT15.

\begin{figure*}
\includegraphics[width = 1.0\hsize]{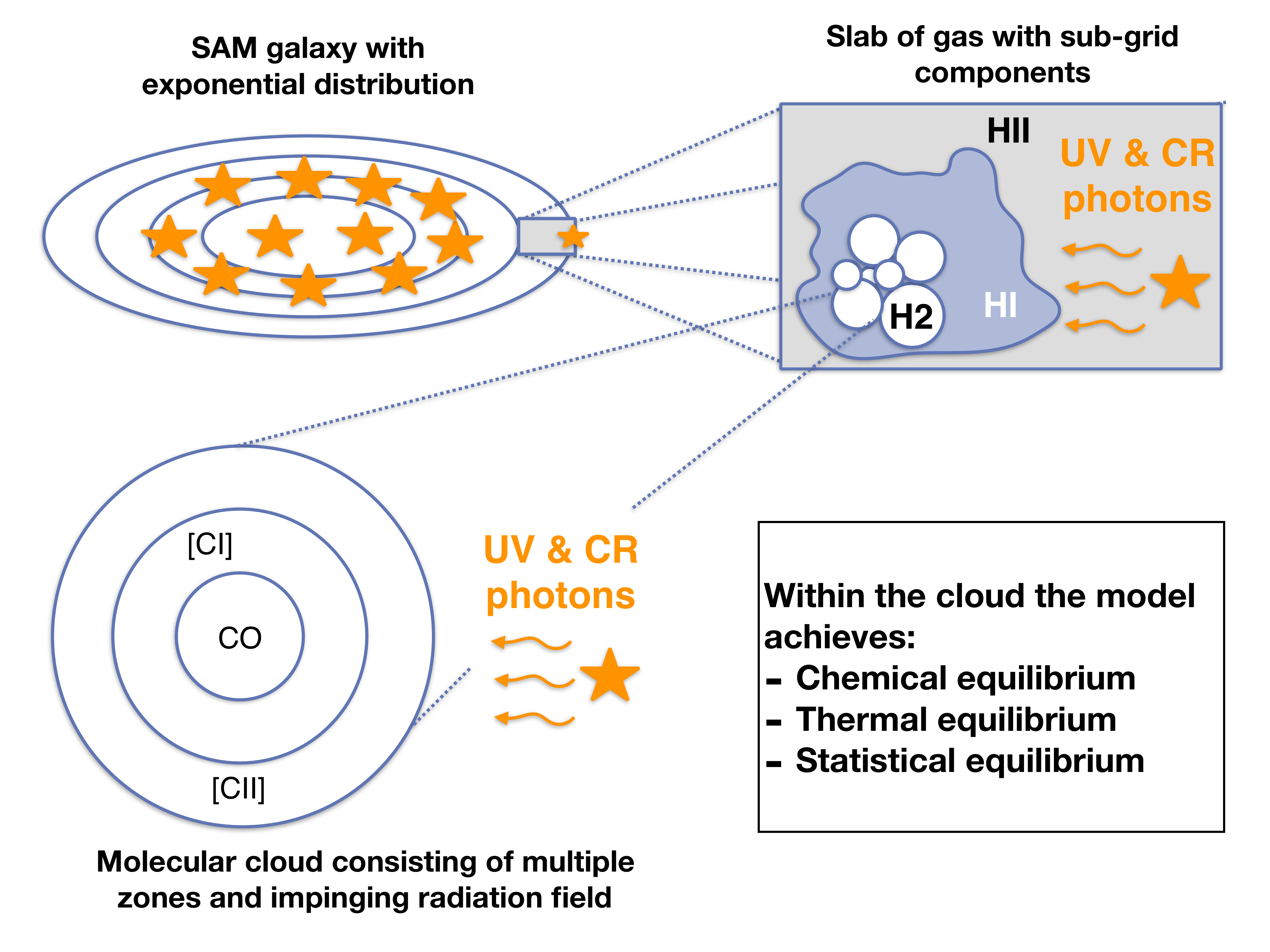}
\caption{A schematic representation of the model presented in this work. Galaxies are represented by an exponential distribution of gas. An annulus of gas within a galaxy consists of ionized, atomic, and molecular gas. The molecular gas is made up by a number of molecular clouds sampled following a molecular cloud mass distribution function. Their sizes are set as a function of the molecular cloud mass and the external pressure acting on the molecular clouds. The individual molecular clouds are made up by radially stratified spheres illuminated by a far-UV (FUV) radiation field and cosmic rays (CR). The molecular clouds are not necessarily assumed to have a fixed average density, but can have a radial density profile. The initial abundance of carbon and oxygen in the ISM is set by the output of the SAM. Within every cloud the model achieves chemical, thermal, and statistical equilibrium. \label{fig:schematic}}
\end{figure*}

\subsection{Sub-mm emission line modeling}
\label{sec:DESPOTIC}
We use \texttt{DESPOTIC} \citep{Krumholz2014} to model the chemistry and sub-mm line emission of individual molecular clouds. This work builds upon the framework described in \citet{Narayanan2017}. We model molecular clouds as radially stratified spheres, where each sphere is chemically and thermally independent from one another. Each cloud contains 25 zones, sufficient to produce converged results for the emergent \CII, \CI, and CO luminosities. We describe the adopted density distribution within the clouds in the following section.

We compute the chemical state of each zone using a reduced carbon-oxygen chemical network \citep{Nelson1999}, in combination with a non-equilibrium hydrogen chemical network \citep{Glover2007,Glover2012}. The chemical reaction and their respective rate coefficients are summarised in Table 2 of 
\citet{Narayanan2017}, and full details on the network are provided in 
\citet{Glover2012}. \texttt{DESPOTIC} requires the strength of the unshielded interstellar radiation field ($G_{\rm UV}$) and the cosmic ray primary ionization rate $\xi_{\rm CR}$ to iterate over the chemical network. The \texttt{DESPOTIC} implementation of the \citet{Glover2012} network includes the effects of dust-shielding on the rates of all photochemical reactions. We describe how $G_{\rm UV}$ and $\xi_{\rm CR}$ are calculated in the following section.

\texttt{DESPOTIC} iteratively solves for the gas and dust temperature and the carbon chemistry within each zone of the molecular clouds. It does this by considering the aforementioned chemical networks and a number of heating and cooling channels. The principal heating processes are heating by the grain photoelectric effect, heating of the dust by the interstellar radiation field, and cosmic ray heating of the gas.  The cooling is dominated by line cooling, as well as cooling of the dust by thermal emission.  Our model also includes cooling by atomic hydrogen excited by electrons via the Lyman $\alpha$ and Lyman $\beta$ lines and the two-photon continuum, using interpolated collisional excitation rate coefficients \citep{Osterbrock2006}. Finally, there is collisional exchange of energy between dust and gas which becomes particularly relevant at relatively high densities ($n\ga 10^4$ cm$^{-3}$).  A full description of these processes is given in \citet{Krumholz2014}.

\texttt{DESPOTIC} solves for the statistical equilibrium within the
level population of each atomic or molecular species. This is done
using the escape probability approximation for the radiative transfer
problem. \texttt{DESPOTIC} accounts for density variations within a zone due to
turbulence, by including a Mach number dependent clumping factor which
represents the ratio between the mass-weighted and volume weighted
density of the gas. It furthermore accounts for the cosmic
microwave background (CMB) as a heating source as well as a background against which emission lines are observed \citep[see for an extensive discussion on the importance of the CMB on sub-mm line emission for example][]{daCunha2013, Vallini2015,Olsen2017, Lagache2017}. We refer the reader to \citet{Krumholz2014} and
\citet{Narayanan2017} for a more detailed description of the \texttt{DESPOTIC} model and the adopted chemical networks. We use the Einstein collisional rate coefficient from the
Leiden Atomic and Molecular Database \citep{Schoeier2005} for our
calculations.

\subsection{Sub-grid physics: coupling the Santa Cruz SAM to \texttt{DESPOTIC}}
In this subsection we describe the different assumptions we make to
couple the Santa Cruz SAM to \texttt{DESPOTIC}. We divide the ISM in
three phases, ionized, atomic, and molecular, as described in Section
\ref{sec:sam}. The density distribution of
  the  ISM in each modeled galaxy follows an exponential
profile. We divide the gas into radial annuli and compute the fraction
of molecular, atomic, and ionized gas as described above. For each
annulus we calculate the sub-mm line emission arising from the
ionized, atomic, and molecular phase. The integrated sub-mm line
emission from a galaxy is calculated by adding the contribution from
each individual annulus. Our sub-grid approaches mostly focus on the
molecular phase, but we will briefly address the atomic and ionized
phases of the ISM towards the end of this Section. A schematic
overview of the coupling between the SAM and \texttt{DESPOTIC} is
depicted in Figure \ref{fig:schematic}. 

We want to emphasize that the sub-resolution models mark an operational prescription to bridge the gap in resolution between galaxy formation models (a SAM in this work) and the small-scale cloud physics. One could think of  alternative prescriptions for the sub-resolution physics than presented in this work. Although interesting, exploring all possible options for each component of the sub-resolution model is a heroic effort too large for a single paper. We rather wish to limit ourselves to a number of well-defined variations in the sub-resolution prescriptions to demonstrate that the resulting sub-mm line emission predicted by models can be highly sensitive to even seemingly minor changes in the sub-resolution physics.

\subsubsection{Molecular cloud distribution function}
The molecular gas within an annulus is made up by a number of individual molecular clouds, the masses $M_{\rm MC}$ of which are assumed to follow a power-law spectrum of the form:
\begin{equation}
\frac{dN}{dM}\propto M^{-\beta},
\end{equation}
where we assume $\beta = 1.8$ based on locally observed cloud
distribution functions
\citep{Solomon87a,Blitz2007,Fukui2008,Gratier2012,Hughes2013,
  Faesi2018}. We will vary this slope in Section \ref{sec:CDF}. We
choose a lower and upper mass limit of $10^4\,\rm{M}_\odot$ and
$10^7\,\rm{M}_\odot$, respectively. For every molecular cloud we
  calculate the total mass of \h2 within it using \texttt{DESPOTIC}
  (the outer regions of a molecular cloud will be ionized/atomic). We
  randomly draw molecular clouds from the distribution function till
  the mass of \h2 within these clouds equals the molecular gas mass as dictated by Equation \ref{eq:H2_GK}.

\subsubsection{Molecular cloud size}
The sizes of molecular clouds $R_{
\rm MC}$ are derived by applying the virial theorem.  $R_{\rm MC}$ depends on the molecular cloud mass and external pressure $P_{\rm ext}$ acting on the molecular cloud \citep{Field2011,Hughes2013,Faesi2018}, such that
\begin{equation}
\frac{R_{\rm MC}}{\rm{pc}} = \biggl(\frac{P_{\rm ext}/k_{\rm B}}{10^4\,\rm{cm}^{-3}\rm{K}}\biggr)^{-1/4}\biggl(\frac{M_{\rm MC}}{290\,\rm{M}_\odot}\biggr)^{1/2},
\end{equation}
where $k_{\rm B}$ is the Boltzmann constant. 

The external pressure at every radius of the galaxy is calculated as a function of the disk mid-plane pressure $P_{\rm m}$. We calculate $P_{\rm m}$ following the approach described in PST14 and SPT15:
\begin{equation}
P_{\rm m}(r) = \frac{\pi}{2}\,G\,\Sigma_{\mathrm{gas}}(r)\left[\Sigma_{\mathrm{gas}}(r) + f_{\sigma}(r)\Sigma_*(r)\right]
\label{eq:pressure}
\end{equation}
where G is the gravitational constant, $f_\sigma(r)$ is the
  ratio between $\sigma_{\mathrm{gas}}(r)$ and $\sigma_*(r)$, the gas
  and stellar vertical velocity dispersion, respectively. The stellar
  surface density profile $\Sigma_*(r)$ is modeled as an exponential
  with scale radius $r_{\mathrm{star}}$ and central density
  $\Sigma_{*, 0} \equiv m_*/(2 \pi r_*^2)$, where $m_*$ is the stellar mass of a galaxy.  Following \citet{Fu2012}, we adopt $f_{\sigma}(r) = 0.1 \sqrt{\Sigma_{*,0}/\Sigma_*}$.

The external pressure $P_{\rm ext}$ is defined as $P_{\rm ext} = P_{\rm m}/(1 + \alpha_0 + \beta_0)$, where $\alpha_0 = 0.4$ and $\beta_0 = 0.25$ account for cosmic and magnetic pressure contributions \citep{Elmegreen1989,Swinbank2011}. The pressure dependence is important, as it partially controls the density of the molecular clouds. In this paper we will explore how the pressure dependence on the size of molecular clouds affects the sub-mm line luminosity of galaxies.

\subsubsection{Density distribution functions within molecular clouds}
\begin{figure}
\includegraphics[width = 1.0\hsize]{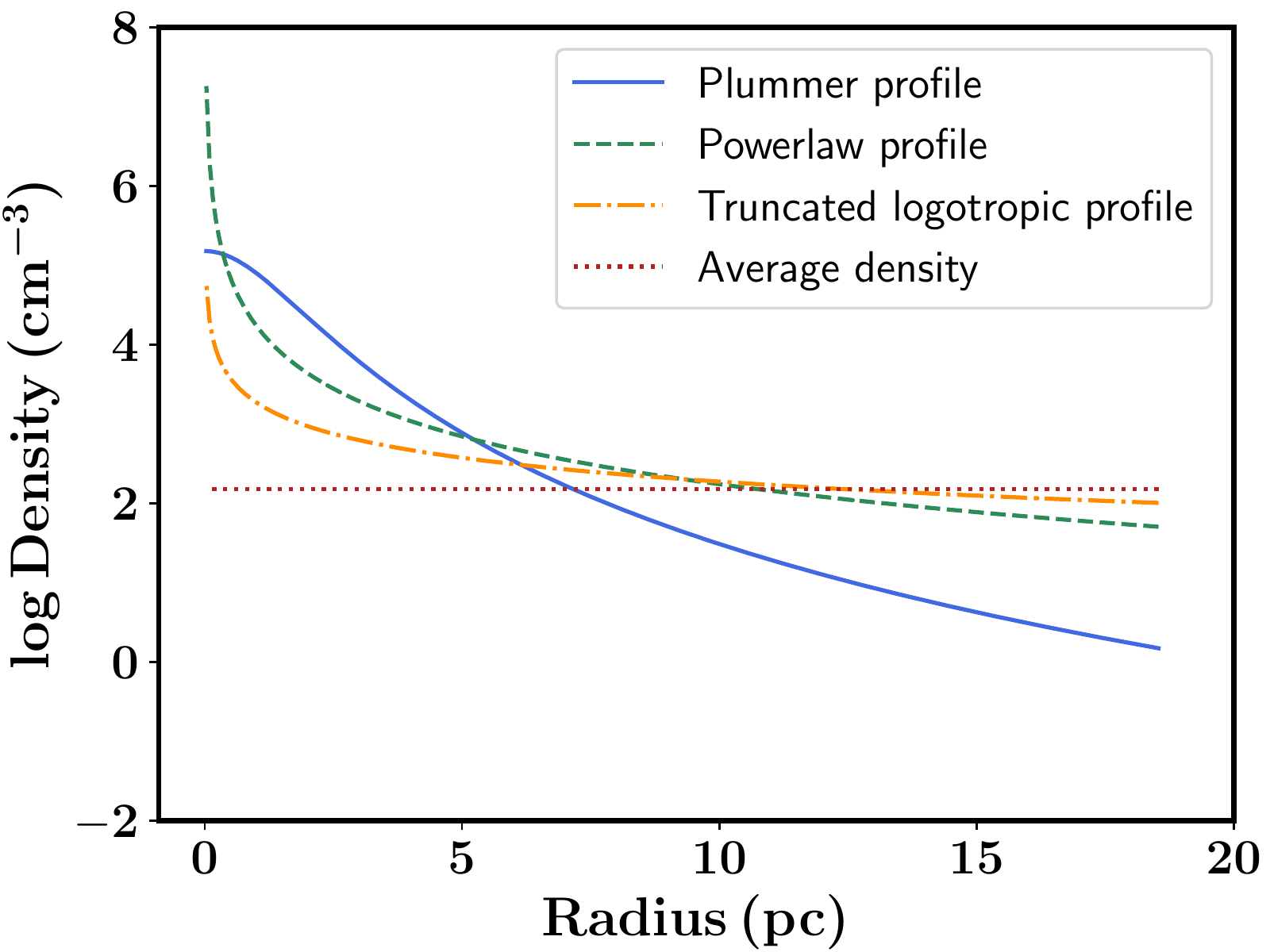}
\caption{A representation of the four different density distribution functions within molecular clouds adopted in this paper. These were obtained assuming a molecular cloud with a mass of $10^5\,\rm{M}_\odot$ and an external pressure acting upon this molecular cloud of $P_{\rm ext}/k_{\rm B} = 10^4\,\rm{cm}^{-3}\rm{K}$. One can clearly see the differences in minimum and maximum densities achieved in the inner and outer regions of the molecular cloud between the different profiles.\label{fig:profiles}}
\end{figure}

We adopt four different approaches to model the density profile of gas
within molecular clouds, a power-law density profile, a Plummer profile, a logotropic density profile, and a fixed average density.  All these four profiles have been adopted in earlier works by different groups and we aim to explore the variation in the predicted sub-mm line luminosities between these density profiles. We describe the different profiles below and an example of each profile is given in Figure \ref{fig:profiles}. It becomes clear that the four different profiles can lead to significant differences in the minimum and maximum densities achieved within a molecular cloud and the radius out to which high density gas (here loosely defined as densities larger than 1000 $\rm{cm}^{-3}$) is present. For all profiles we take $n_{\rm H}(R > R_{\rm MC}) = 0\,\rm{cm}^{-3}$. It can be expected that in reality individual giant molecular clouds follow a more complex hierarchichal density structure. The four adopted profiles thus mark an operational definition for the density distribution within molecular clouds, (note that on top of this we account for turbulence driven variations in the densities as explained in Section \ref{sec:DESPOTIC}). They should therefore be thought of as physically inspired, but not literal density distributions.

\subsubsection*{Power-law profile}
The molecular clouds are modeled as a power-law sphere  where the density is given by
\begin{equation}
n_{\rm H}(R) = n_0\biggl(\frac{R_{\rm MC}}{R}\biggr)^{-\alpha},
\end{equation}
where $\alpha$ is set to $\alpha = 2$ \citep{Walker1990}. 

\subsubsection*{Plummer profile}
The Plummer profile assures a finite central density and was suggested by \citet{Whitworth2001} to fit the observed density profiles of prestellar cores and class 0 protostars. This profile was also adopted by \citet{Olsen2015CO}. The radial density profile is described as:
\begin{equation}
n_{\rm H}(R) = \frac{3M_{\rm MC}}{4\pi R_{\rm p}^3}\biggl(1 + \frac{R^2}{R_{\rm p}^2}\biggr)^{-5/2},
\end{equation}
where $R_{\rm p}$ is the Plummer radius, which is set to  $R_{\rm p} = 0.1 R_{\rm MC}$ following \citet{Olsen2015CO}. 

\subsubsection*{Logotropic profile}
The radial density profiles of the molecular clouds are assumed to follow a truncated logotropic profile \citep{Olsen2017},
\begin{equation}
n_{\rm H}(R) = n_{\rm H,ext}\frac{R_{\rm MC}}{R},
\end{equation}
where the external density $n_{\rm H,ext}$ is two-thirds of the average density within $R_{\rm MC}$.

\subsubsection*{Fixed average density}
The molecular clouds have a uniform density (i.e., a flat density profile) derived from their mass $M_{\rm MC}$ and size $R_{\rm MC}$.

\subsubsection{Impinging UV radiation field and cosmic ray strength}
We scale the strength of the UV radiation field $G_{\rm UV}$ directly with the local SFR surface density $\Sigma_{\rm SFR}$:
\begin{equation}
G_{\rm UV} = G_{\rm UV,MW} \times \frac{\Sigma_{\rm SFR}}{\Sigma_{\rm SFR,MW}},
\end{equation}
where $G_{\rm UV}$ and $G_{\rm MW, UV}$ are expressed in Habing units and $G_{\rm MW, UV} = 9.6 \times 10^{-4}\,\rm{erg}\,\rm{cm}^{-2}\,\rm{s} = 0.6 
\,\rm{Habing}$ \citep{Seon2011} and $\Sigma_{\rm SFR,MW} = 0.001\,\rm{M}_\odot$ \citep{Bonatto2011}. The cosmic ray field $\xi_{\rm CR}$ is also scaled as a function of the local SFR surface density such that
\begin{equation}
\label{eq:CR_scaling}
\xi_{\rm CR} = 0.1\,\xi_{\rm CR,MW} \times \frac{\Sigma_{\rm SFR}}{\Sigma_{\rm SFR,MW}},
\end{equation}
where $\xi_{\rm CR,MW} = 10^{-16}\rm{s}^{-1}$ following \citet{Narayanan2017}.

\subsubsection{Elemental abundances}
The elemental abundance of carbon [C/H] and oxygen [O/H] are scaled as
a function of the gas phase metallicity of the cold gas $Z_c$ as
predicted by the SAM, such that [C/H] $= Z_c \times 2\times 10^4$ and
[O/H] $= Z_c \times 4\times 10^4$ \citep{Draine2011}.

\subsubsection{Contribution from the atomic diffuse ISM}
Besides the molecular ISM, the atomic diffuse ISM may also contribute to the \CII emission of galaxies. To include the contribution from this ISM phase we model the atomic diffuse ISM as one-zone clouds. These clouds are illuminated by a UV radiation field and cosmic-ray field strength scaled by the integrated SFR of the galaxy normalised by a SFR of $1\,\rm{M}_\odot\,\rm{yr}^{-1}$ ($G_{\rm UV} = G_{\rm UV,MW} \times \rm{SFR}$ and $\xi_{\rm CR} = 0.1\,\xi_{\rm CR,MW}\times\rm{SFR}$). These one-zone clouds have a column density of $N_{\rm H} = 10\times 10^{20}\,\rm{cm}^{-2}$ and a hydrogen density of $n_H = 10\,\rm{cm}^{-3}$  \citep{Elmegreen1987,McKee2015}. 
The \CII, \CI, and CO line-emission contribution by the atomic diffuse gas is added to the contribution by the molecular gas.

\begin{figure}
\includegraphics[width = 1.0\hsize]{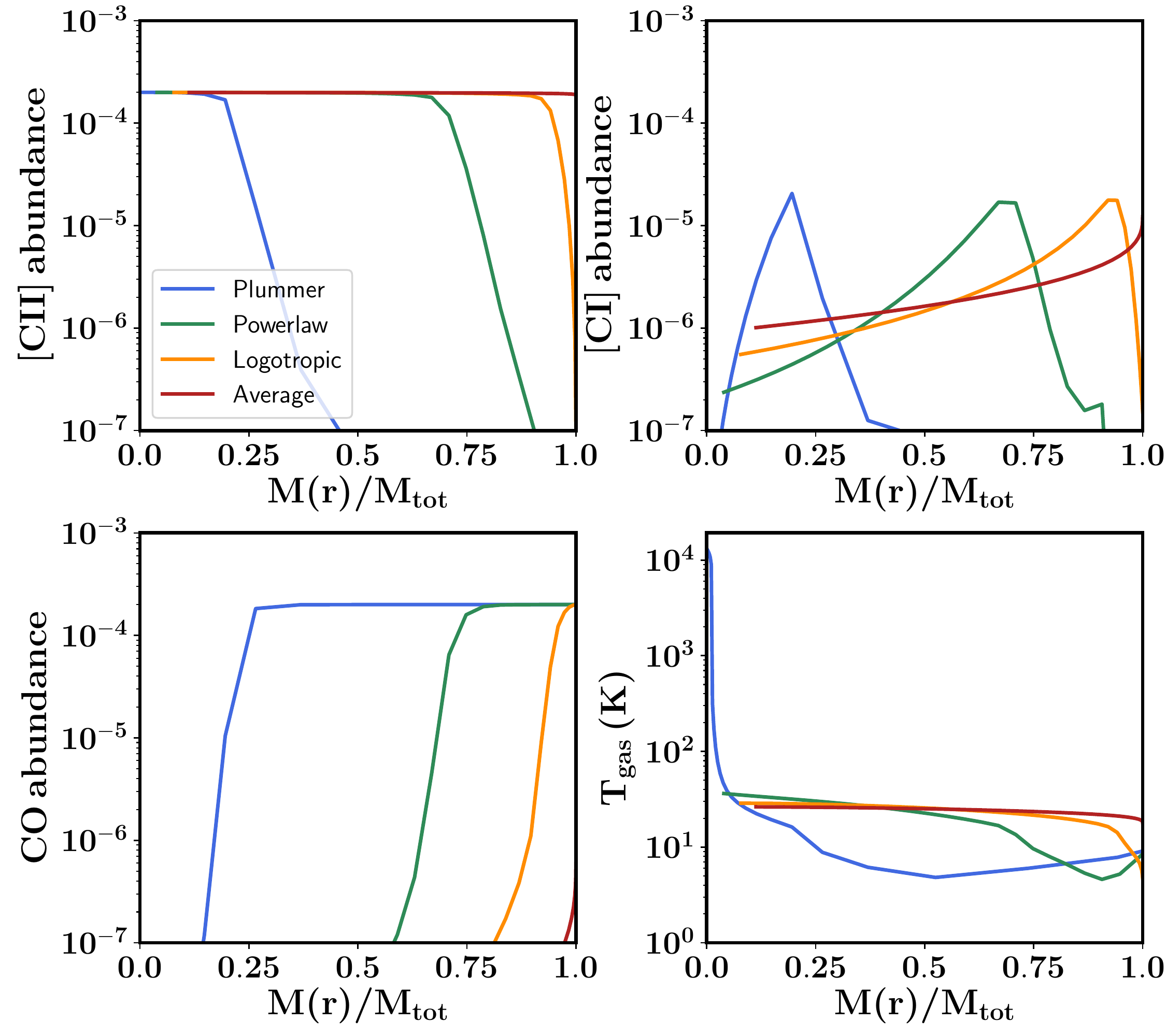}
\caption{The \CII (top left), 
\CI (top right), and CO (bottom left) abundance and gas temperature
(bottom right) profiles of a molecular cloud for different molecular
cloud density profiles. The molecular cloud has a fixed mass of $10^5\,\rm{M}_\odot$,  an external pressure acting upon it of $P_{\rm ext}/k_{\rm B} =
10^4\,\rm{cm}^{-3}\,\rm{K}$, a UV radiation field shining on it of one
$G_0$, and a solar metallicity. The different density profiles lead to very
different radial profiles for the CO, \CI, and \CII abundance and
temperature of the gas.}\label{fig:profiles_abund}
\end{figure}

\begin{figure}
\includegraphics[width = 1.0\hsize]{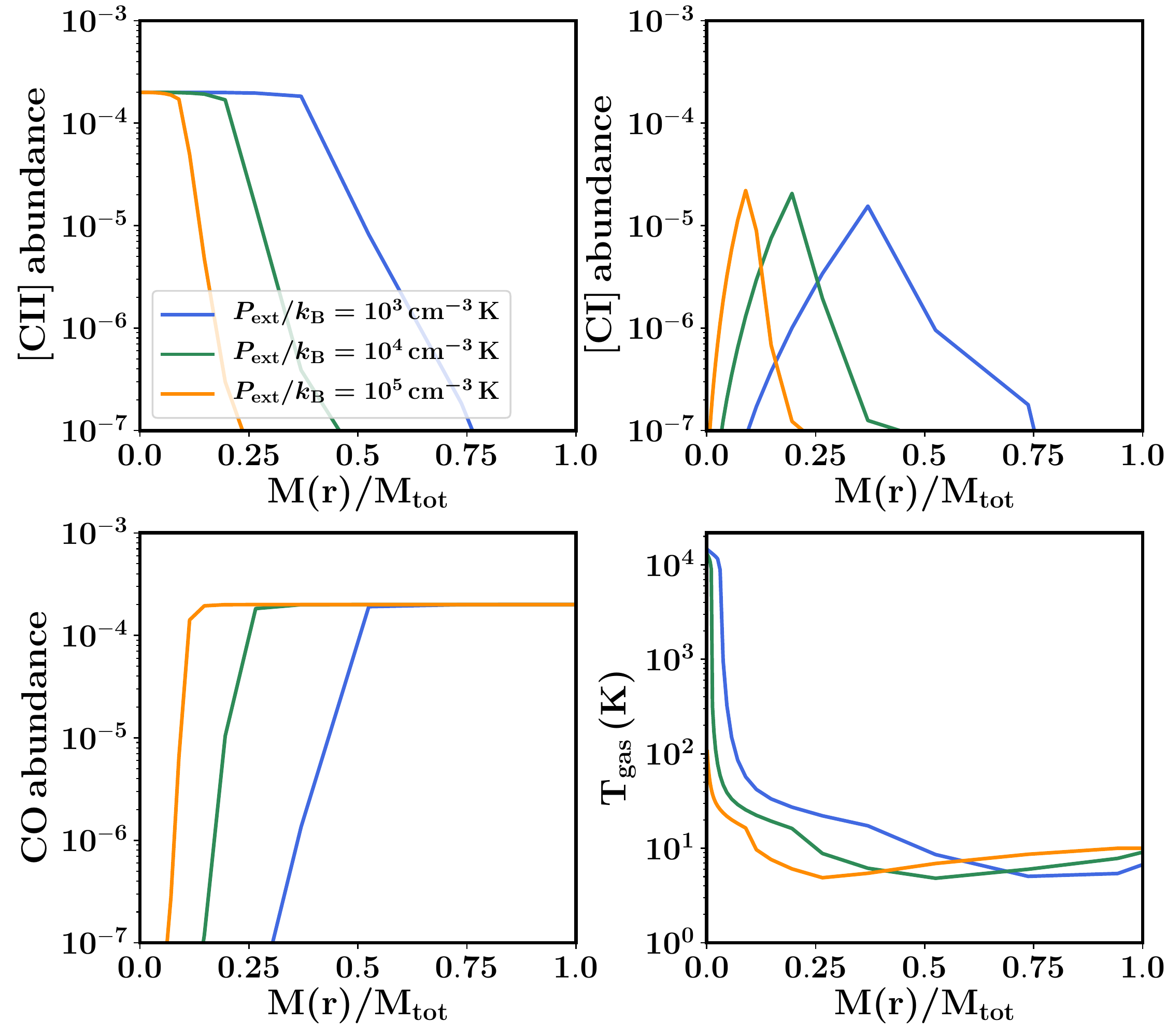}
\caption{The \CII (top left), 
\CI (top right), and CO (bottom left) abundance and gas temperature
(bottom right) profiles of a molecular cloud while varying the
external pressure acting upon the molecular cloud. The molecular cloud
has a fixed mass of $10^5\,\rm{M}_\odot$ distributed following a
plummer density profile,  a UV radiation field shining on it of one
$G_0$, and a solar metallicity. As the external pressure increases,
the CO abundances increases, whereas the \CII abundance decreases. The
gas temperatures within the molecular cloud also decrease with
increasing external pressure.}\label{fig:pressure_abund}
\end{figure}

\begin{figure}
\includegraphics[width = 1.0\hsize]{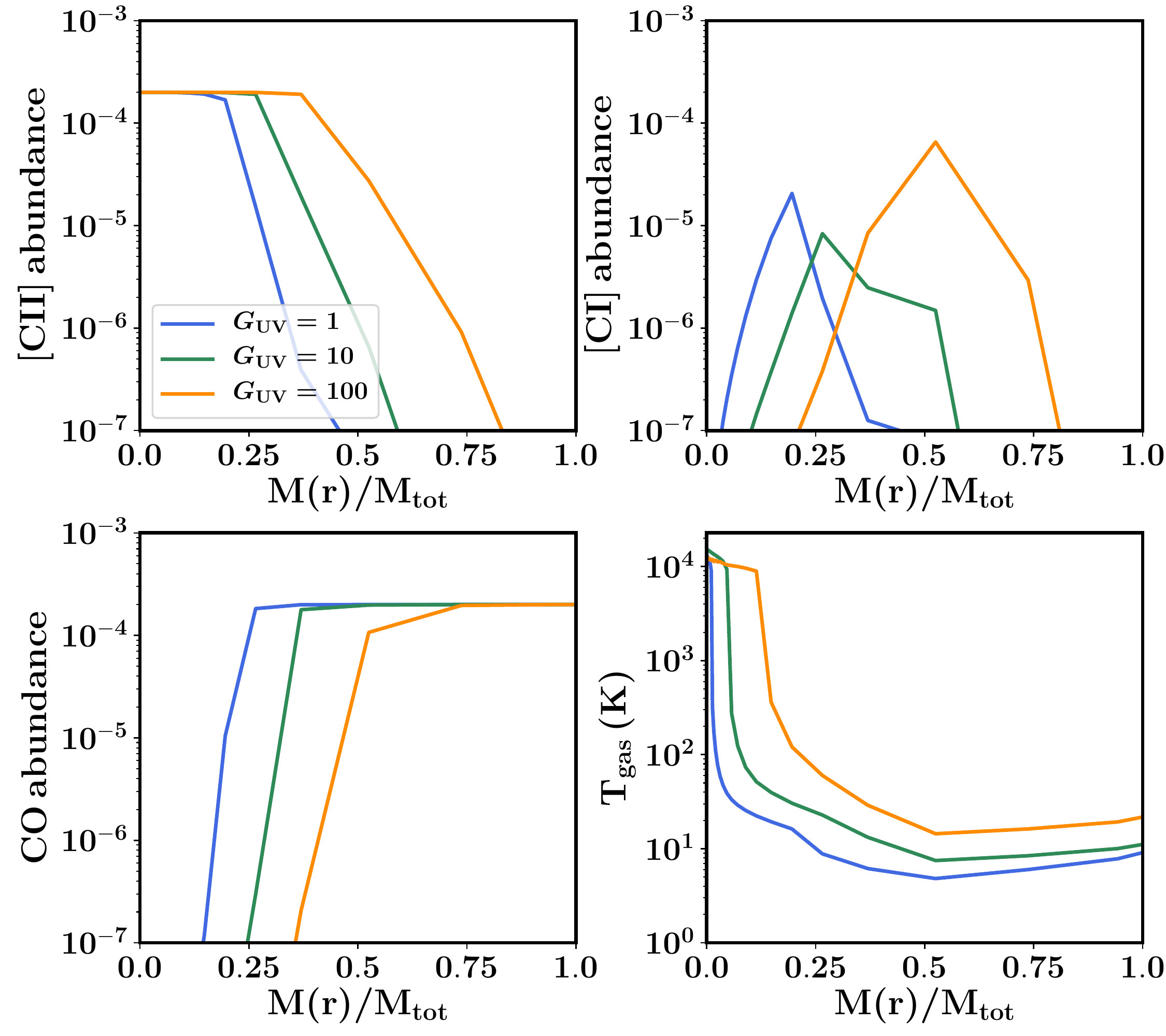}
\caption{The \CII (top left), 
\CI (top right), and CO (bottom left) abundance and gas temperature
(bottom right) profiles of a molecular cloud for different strengths
of impinging UV radiation field. The molecular cloud has a fixed mass
of $10^5\,\rm{M}_\odot$ distributed following a plummer density profile, an external pressure acting upon it of $P_{\rm ext}/k_{\rm B} =
10^4\,\rm{cm}^{-3}\,\rm{K}$, and a solar metallicity. As the strength of the UV radiation field increases, the \CII abundance and gas temperature become higher, whereas the \CI and CO abundances are lower. Especially the temperature reacts very strongly on the strength of the UV radiation field, particularly in the regime where \CII dominates.}\label{fig:radfield_abund}
\end{figure}

\begin{figure*}
\includegraphics[width = 1.0\hsize]{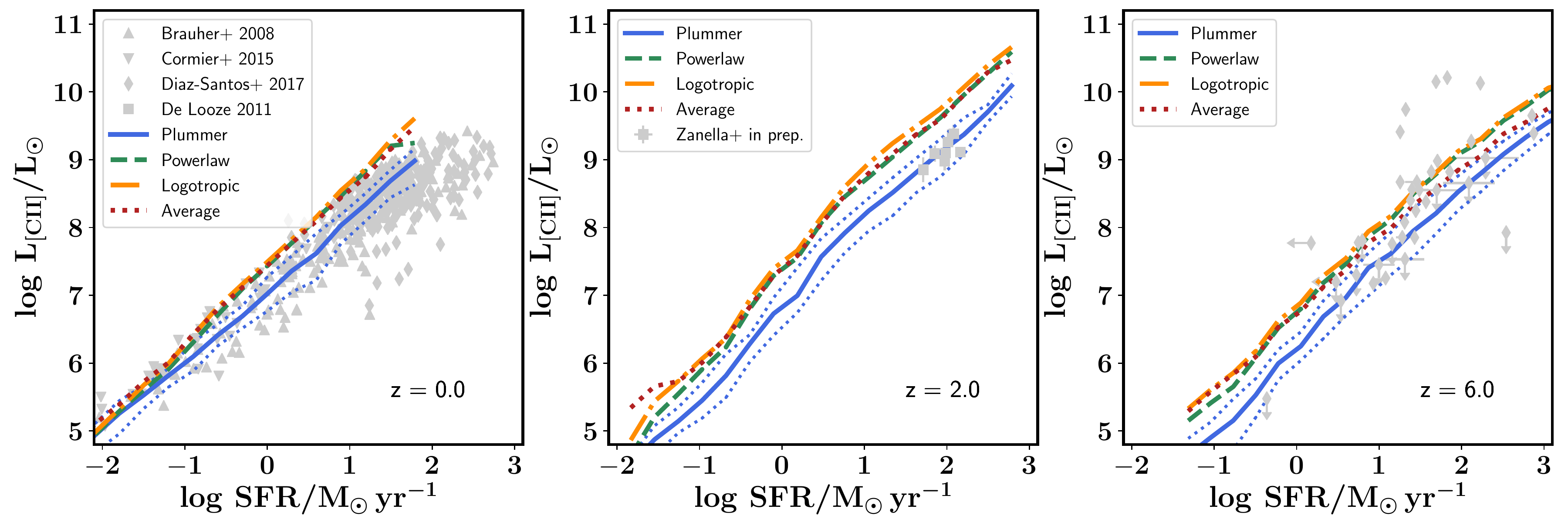}
\caption{The \CII luminosity of galaxies as a function of their SFR at $z=0$, $z=2$, and $z=6$, assuming different radial
  density profiles for the gas within molecular clouds. Model
  predictions are compared to observational constraints \citep[and
  Zanella et al. in
  prep]{Brauher2008,deLooze2011, Cormier2015,DiazSantos2017,Capak2015,Knudsen2016,Willott2015,Decarli2017,GonzalezLopez2014,Kanekar2013,Pentericci2016,Bradac2017,Schaerer2015,Maiolino2015,Ota2014,Inoue2016,Knudsen2017,Carniani2017}. In
  this particular plot the Plummer model represents our fiducial
  model. Changing the density profile of molecular clouds can lead to
  variations up to $\sim$0.5 dex in the predicted \CII luminosity of
  actively star-forming galaxies.\label{fig:CIIdensity}}
\end{figure*}

\section{Carbon Chemistry}
\label{sec:carbon_chemistry}
Before presenting the CO, \CI, and \CII luminosity of galaxies when
varying between different sub-grid recipes, we first explore how these
choices affect the carbon chemistry \citep[similar exercises have been performed before in e.g.,][]{Wolfire2010,Bisbas2015,Bisbas2017}.

In Figure
\ref{fig:profiles_abund} we show the CO, \CI, and \CII abundance
profile of a molecular cloud, as well as its temperature profile, when varying
the density profile within the molecular cloud. For all these
scenarios we assume a molecular cloud with a fixed mass of $10^5\,\rm{M}_\odot$,  an external pressure acting upon of $P_{\rm ext}/k_{\rm B} =
10^4\,\rm{cm}^{-3}\,\rm{K}$, a UV radiation field of 1 $G_0$, and a
solar metallicity  (at $z=0$). We find that the different density profiles result
in very different CO, \CI, and \CII abundance and temperature
profiles. The Plummer density profile results in the largest mass fraction of
CO, whereas adopting the fixed average density profile results in  hardly
any CO. The radius at which the \CI abundance dominates varies
significantly between the different density profiles. The gas
temperature distribution is also very different between the different
profiles. The gas temperature is
highest at the edge of the molecular clouds when adopting the Plummer profile, but quickly drops to
temperatures of $\sim 10\,\rm{K}$.\footnote{ We note that the CMB sets a floor for the temperature of the molecular clouds which is above 10 K already at $z=2.7$.}  For the other profiles we find a
temperature of $\sim 30\,\rm{K}$ over a large fraction of the
molecular cloud with a drop in temperature further inwards of the
molecular clouds. Overall we find that the Plummer profile predicts
much higher CO abundances and lower gas temperatures.  The  reason for this is that the Plummer profile has a long tail
towards larger radii with relatively high densities (a few 1000
$\rm{cm}^{-3}$, see Figure \ref{fig:profiles}). This tail constitutes
a large mass fraction and contributes significantly to the overall CO
abundance and allows for efficient cooling of the gas.

In Figure \ref{fig:pressure_abund} we show the CO, \CI, and \CII
abundances of a molecular cloud when changing the external pressure
acting upon the molecular cloud (molecular cloud properties are otherwise
similar as in Figure \ref{fig:profiles_abund}, assuming a plummer
density profile). As the pressure acting upon the molecular cloud
increases, the density of the molecular cloud increases as well. As a
result, a higher fraction of the carbon is locked up in CO, whereas
the \CII abundance rapidly decreases. The increased density furthermore leads to a decrease in the gas temperature as a function of external pressure.

In Figure \ref{fig:radfield_abund} we show the CO, \CI, and \CII
abundances of a molecular cloud when changing the UV radiation field
(molecular cloud properties are otherwise
similar as in Figure \ref{fig:profiles_abund}, assuming a plummer
density profile).  An increase in the UV radiation field  results in a more effective dissociation of the CO molecules \citep[e.g.,][]{Hollenbach1991,Wolfire2010}, which lowers the CO abundance. Furthermore, the \CII
abundance increases and the gas
temperature increases.

Our results are in agreement with the findings by \citet{Wolfire2010} and \citet{Bisbas2015}. For example, these authors also find that when the UV and/or cosmic ray field increases, the CO is more centrally concentrated within a molecular cloud.

\section{CO, \CI, and \CII luminosities of galaxies}
\label{sec:results}
In this section we present our predictions for the CO, \CI, and \CII
emission of galaxies, while varying the sub-grid components of our
model. We restrict our analysis to central star forming galaxies,
selected using the criterion $\rm{sSFR} > 1/(3t_H(z))$, where
$\rm{sSFR}$ is the galaxy specific star-formation rate and $t_H(z)$
the Hubble time at the galaxy's redshift. This approach selects
galaxies in a similar manner to commonly used observational methods
for selecting star-forming galaxies, such as color-color cuts
\citep[e.g.,][]{Lang2014}. We present the 14th, 50th, and 86th
percentile of the different model variants in every figure. The 50th
percentile corresponds to the median, the 14th percentile corresponds
to the line below which 14 per cent of the galaxies are located,
whereas the 86th percentile corresponds to the line below which 86 per
cent of the galaxies are located. We typically only show the 14th and
86th for one model variant to increase the clarity of the figures. The
scatter is always similar between the different model variants.
  
Throughout the rest of the paper we will present our model predictions
in four different plots, focusing on the \CII, \CI, and CO emission of
galaxies. \CII comparisons between model predictions and observations
are performed using data presented in \citet{Brauher2008}, \citet{deLooze2011}, \citet{Cormier2015}, \citet{DiazSantos2017} at
$z=0$, Zanella et al. (in prep.) at $z=2$, and a compilation of
observations at $z\sim6$ \citep{Capak2015,Knudsen2016,Willott2015,
  Decarli2017,GonzalezLopez2014,Kanekar2013,Pentericci2016,Bradac2017,Schaerer2015,Maiolino2015,Ota2014,Inoue2016,Knudsen2017,Carniani2017}. The
comparison for \CI is performed using $z=0$ observations by
\citet{Gerin2000}. CO comparisons are carried out using data presented
in \citet{Leroy2008}, \citet{Papadopoulos2012}, \citet{Greve2014},
\citet{Kamenetzky2016}, \citet{Liu2015}, \citet{Cicone2017}, and
\citet{Saintonge2017} for $z=0$, and \citet{Tacconi2010} and
\citet{Tacconi2013} for $z=1$ and $z=2$. Infrared luminosities from the
literature were converted into star-formation rates following the
infrared--SFR relation in \citet[comes from \citet{Murphy2011}]{Kennicutt2012}.

In some cases the differences between the predictions by different sub-grid model variants are very minimal and are shown in the Appendix rather than the main body of this paper.

\begin{figure*}
\includegraphics[width = 1.0\hsize]{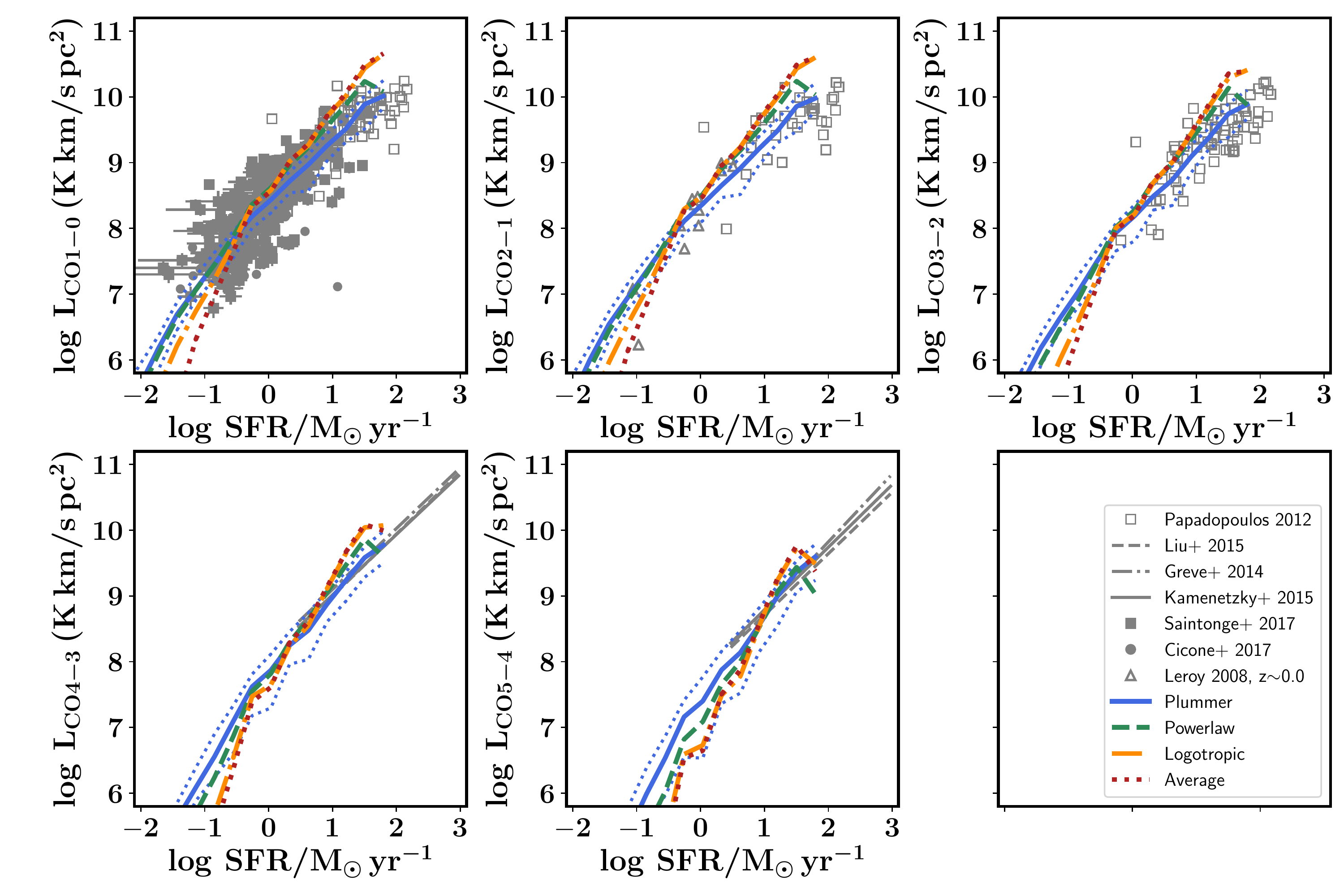}
\caption{The CO J$=$1--0 to CO J$=$5--4 luminosity of galaxies as a
  function of their SFR at $z=0$ assuming different radial density
  profiles for the gas within molecular clouds. Model predictions are
  compared to observational constraints taken from \citet{Leroy2008},
  \citet{Papadopoulos2012}, \citet{Cicone2017},
  \citet{Saintonge2017}, \citet{Greve2014},
\citet{Kamenetzky2016}, and \citet{Liu2015}.  In
  this particular plot the Plummer model represents our fiducial
  model. Changing the density profile of molecular clouds can lead to
  variations up to $\sim$0.5 dex in the predicted CO luminosity of galaxies.\label{fig:COz0density}}
\end{figure*}

\begin{figure*}
\includegraphics[width = 1.0\hsize]{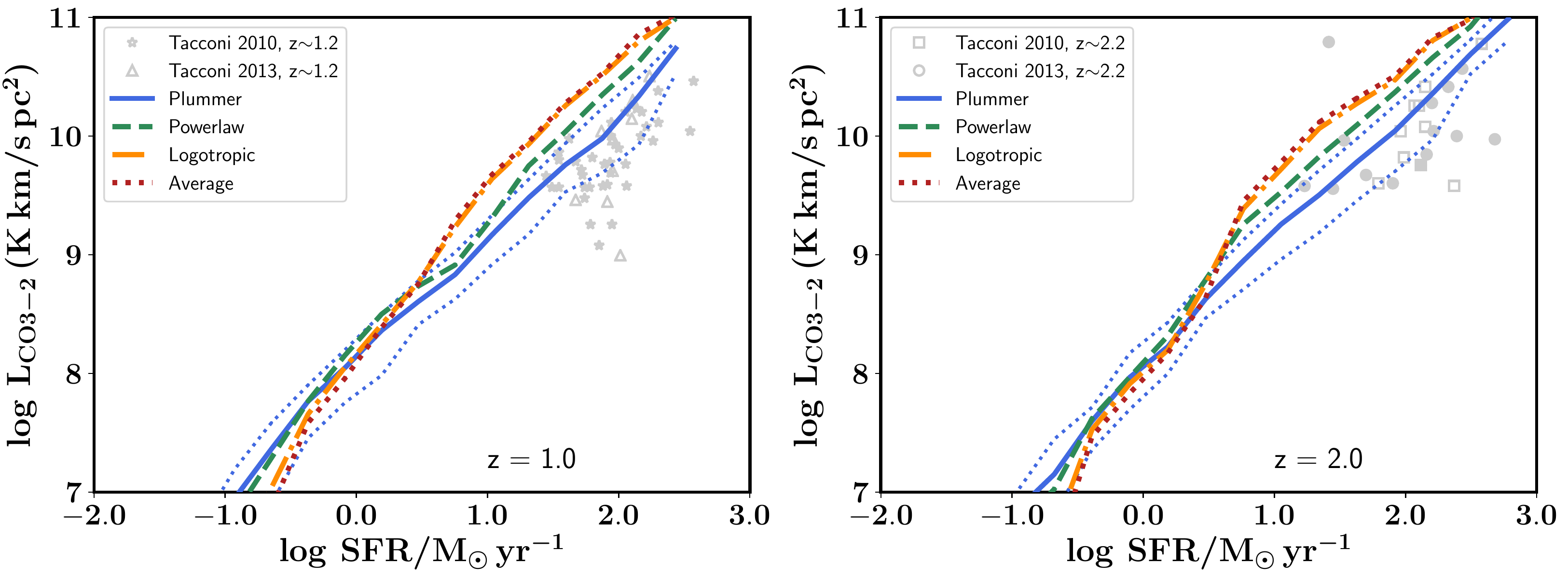}
\caption{The CO J$=$3--2 luminosity of galaxies at $z=1$ and $z=2$ as a function of their SFR assuming different radial density profiles for the gas within molecular clouds. Model predictions are compared to observational constraints taken from \citet{Tacconi2010}, and \citet{Tacconi2013}.  In
  this particular plot the Plummer model represents our fiducial
  model. Changing the density profile of molecular clouds can lead to
  variations up to $\sim$0.5 dex in the predicted CO luminosity of
  galaxies. The Powerlaw, Logotropic, and Average density profiles predict CO luminosities that are too bright in $z=1$ and $z=2$ galaxies.\label{fig:COzhighdensity}}
\end{figure*} 

\begin{figure}
\includegraphics[width = 1.0\hsize]{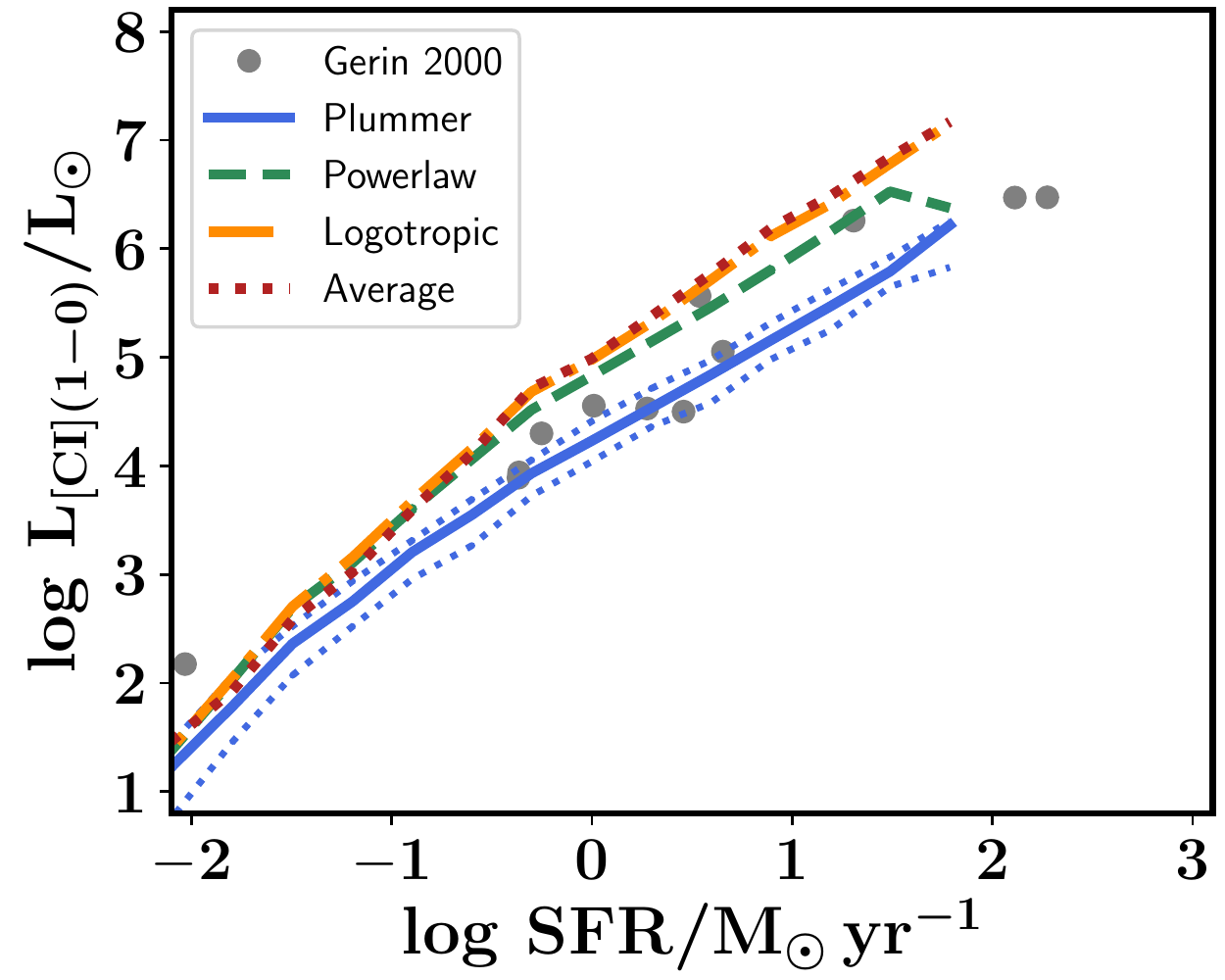}
\caption{The [CI] 1--0 luminosity of galaxies at $z=0$ as a function
  of their SFR assuming different radial density profiles for the gas within molecular clouds. Model predictions are compared to observational constraints taken from \citet{Gerin2000}.  In
  this particular plot the Plummer model represents our fiducial
  model. Changing the density profile of molecular clouds can lead to
  variations up to one dex in the predicted \CI luminosity of galaxies.\label{fig:CIdensities}}
\end{figure}

\subsection{Varying density profiles}
In Figure \ref{fig:CIIdensity} we present model predictions for the
\CII luminosity of galaxies as a function of their SFR at
$z=0$, $z=2$, and $z=6$. We show this
for the four molecular cloud density profiles discussed in this
work. We find that three of the four density profiles (Power-law,
Logotropic, and Average) predict almost identical \CII luminosities
for galaxies at all redshifts considered.  The Plummer density profile
predicts \CII luminosities that are approximately 0.5 dex lower than
the other profiles, independent of redshift. 
  The luminosities predicted by the Powerlaw,
Logotropic, and Average density profiles are too high compared to the observations at $z=0$, $z=2$ at $z=6$ \citep[except for a handful
  galaxies with a SFR of 10--100 $\rm{M}_\odot\,\rm{yr}^{-1}$ and \CII
  luminosity brighter than $10^{10}\,\rm{L}_\odot$ from
][Note however that \citet{Faisst2017} suggests that the estimated
SFRs of the Capak et al. sources are too low.]{Capak2015}. Overall, the model adopting the Plummer profile does
best at reproducing the \CII luminosity of galaxies from $z=0$ to $z=6$. The fainter \CII luminosities predicted by the Plummer profile are driven by lower \CII abundances throughout most of the molecular cloud compared to the other density profiles (see Figure \ref{fig:profiles_abund}.

In Figure \ref{fig:COz0density} we show the predicted CO J$=$1--0, CO
J$=$2--1, CO J$=$3--2, CO J$=$4--3, and CO J$=$5--4 luminosities of
galaxies at $z=0$ as a function of their SFR. Here again,
the Plummer profile predicts luminosities significantly lower than the
other three density profiles, up to almost an order of magnitude
towards the most actively star-forming galaxies for all CO rotational
transitions. The Logotropic and Average profiles predict CO
luminosities that are a bit brighter than the Powerlaw
profile. The Plummer profile predicts CO luminosities brighter than the other profiles for galaxies with a SFR less than $1\,\rm{M}_\odot\,\rm{yr}^{-1}$. Overall the Plummer profile best reproduces the CO J$=$1--0 through CO J$=$5--4 luminosity of local galaxies over a large range in
SFR. We find similar differences between the four density
profiles when looking at the CO luminosities of $z=1$ and $z=2$
galaxies as a function of their SFR (Figure
\ref{fig:COzhighdensity}). The Plummer density profile reproduces the
CO luminosities of $z=1$ and $z=2$ galaxies best, whereas the other
profiles predict CO luminosities $\sim 0.3$ dex higher. The brighter CO emission predicted by the Plummer profile in galaxies with a SFR less than $1\,\rm{M}_\odot\,\rm{yr}^{-1}$ is caused by the broad wing of the Plummer profile. This is clear in Figure \ref{fig:profiles}, where we see that for a cloud with a mass of $10^5\,\rm{M}_\odot$ the Plummer profile predicts the highest densities from 1 to 5 pc. In Figure \ref{fig:profiles_abund} we then see that this indeed causes a higher CO abundance for a large fraction of the molecular clouds. This contribution makes a big difference in galaxies with low SFRs, which in the SAM are galaxies with lower gas surface densities and hence lower average volume densities.

We present the \CI 1--0 luminosity of galaxies at $z=0$ as a function
of their SFR in Figure \ref{fig:CIdensities}. There is only
little difference between the Powerlaw, Logotropic, and Average model
variants. The Plummer profile predicts \CI 1--0 luminosities that are
almost an order of magnitude fainter than the other model variants.  Best agreement with the observations is found for the Plummer profile model variants.

\begin{figure*}
\includegraphics[width = 1.0\hsize]{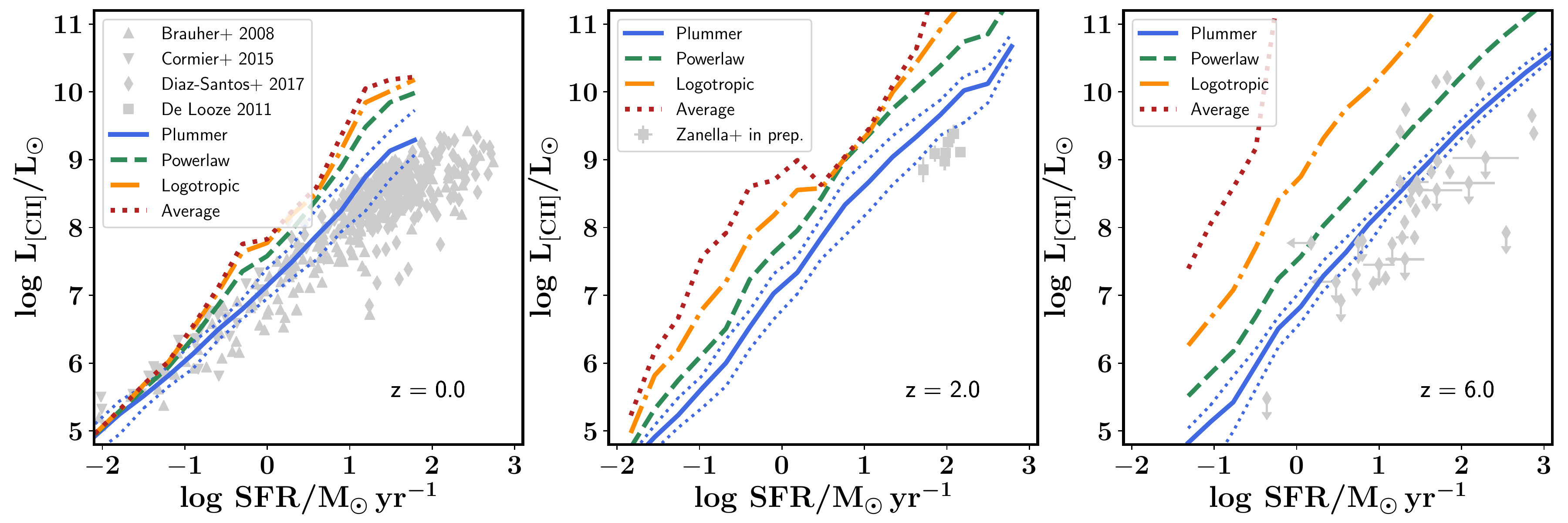}
\caption{The \CII luminosity of galaxies as a function of their SFR at $z=0$, $z=2$, and $z=6$ for different radial density
  profiles for the gas within molecular clouds and assuming a fixed
  external pressure acting on the molecular clouds of $P_{\rm
    ext}/k_{\rm B} = 10^4\,\rm{cm}^{-3}\,\rm{K}$. This Figure is
  similar to Figure \ref{fig:CIIdensity}.\label{fig:CIInopressure},
  aside from the fact that here we impose a constant external
    pressure on clouds.  When imposing a constant external pressure on the cloud the predicted \CII luminosities increase. }
\end{figure*}

\begin{figure*}
\includegraphics[width = 1.0\hsize]{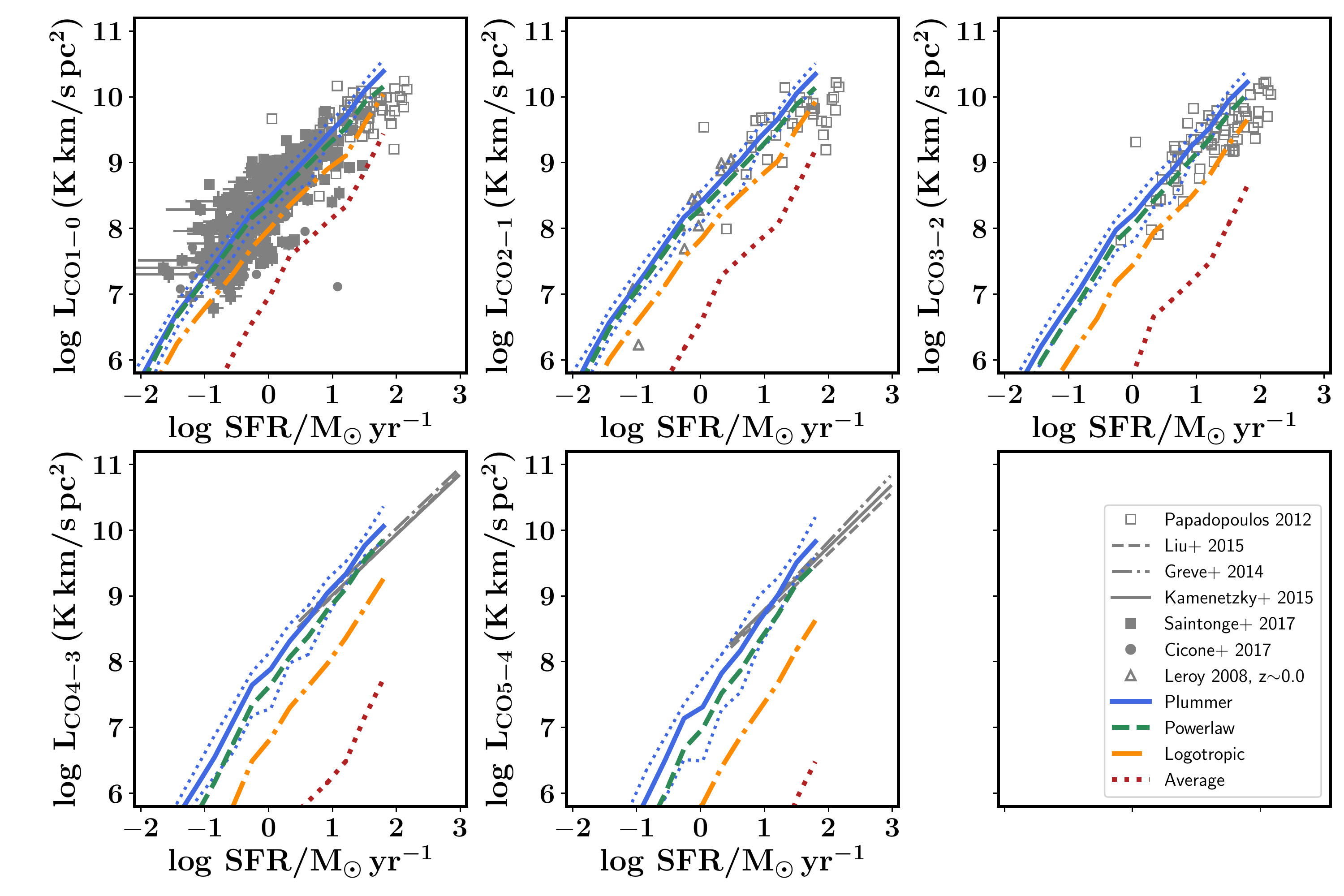}
\caption{The CO J$=$1--0 to CO J$=$5-4 luminosity of galaxies as a
  function of their SFR at $z=0$ assuming different radial density
  profiles for the gas within molecular clouds and a fixed external
  pressure acting on the molecular clouds of $P_{\rm ext}/k_{\rm B} =
  10^4\,\rm{cm}^{-3}\,\rm{K}$. This Figure is similar to Figure
  \ref{fig:COz0density}. When imposing a constant external pressure on
  the cloud the predicted CO luminosities decrease. This decrease is
  most dramatic and in strong tension with the observations for the
  Logotropic and Average density profiles. The CO luminosities
  predicted by the Plummer model variant are a little bit brighter.\label{fig:COz0pressure}}
\end{figure*}

\begin{figure*}
\includegraphics[width = 1.0\hsize]{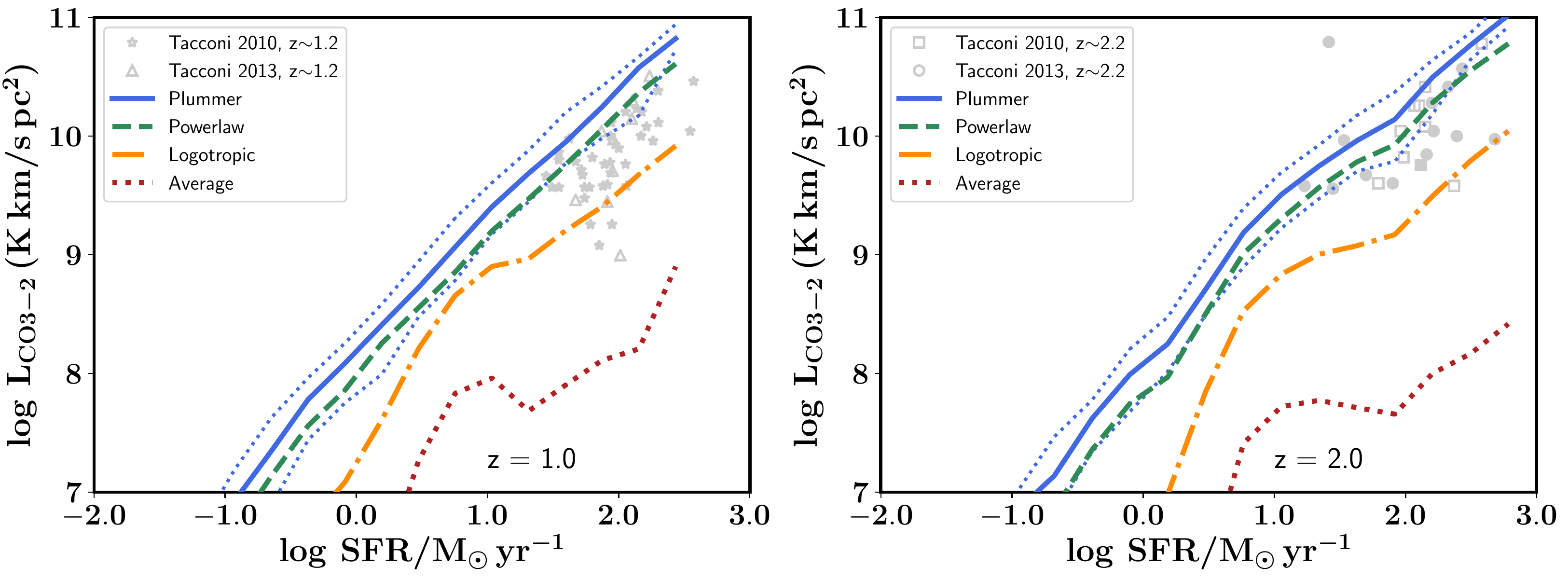}
\caption{The CO J$=$3--2 luminosity of galaxies at $z=1$ and $z=2$ as
  a function of their SFR assuming different radial density profiles
  for the gas within molecular clouds and a fixed external pressure
  acting on the molecular clouds of $P_{\rm ext}/k_{\rm B} =
  10^4\,\rm{cm}^{-3}\,\rm{K}$. This Figure is similar to Figure
  \ref{fig:COzhighdensity}. When imposing a constant external pressure
  on the cloud the predicted CO luminosities decrease. This decrease
  is most dramatic and in strong tension with the observations for the
  Logotropic and Average density profiles. The CO luminosities
  predicted by the Plummer model variant are a little bit brighter.\label{fig:COhighzPressure}}
\end{figure*}

\begin{figure}
\includegraphics[width = 1.0\hsize]{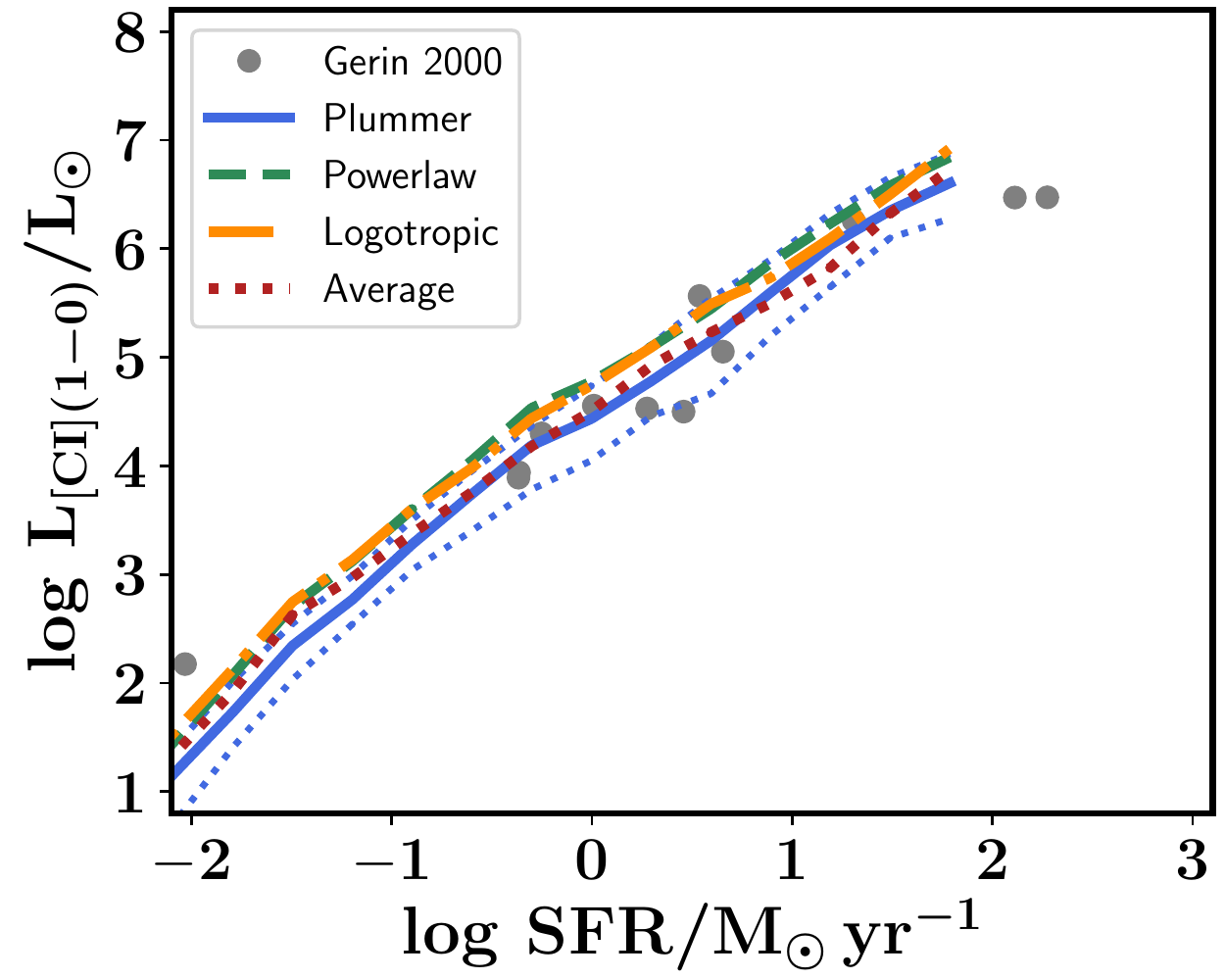}
\caption{The [CI] 1--0 luminosity of galaxies at $z=0$ as a function of their SFR assuming different radial density profiles for the gas within molecular clouds a fixed external pressure acting on the molecular clouds of $P_{\rm ext}/k_{\rm B} = 10^4\,\rm{cm}^{-3}\,\rm{K}$. This Figure is similar to Figure \ref{fig:CIdensities}. When imposing a constant external pressure on the cloud the \CI luminosities predicted by the various density profiles are almost identical.\label{fig:CIpressure}}
\end{figure}

\subsection{No pressure acting on molecular clouds}
\label{sec:pressure}
In this subsection we explore the importance of the pressure
dependence of the molecular cloud size for the sub-mm line luminosity
of galaxies. In Figure \ref{fig:CIInopressure} we show the \CII
luminosity of galaxies as a function of their SFR where we
assume the external pressure is a constant $P_{\rm ext}/k_{\rm
  B} = 10^4\,\rm{cm}^{-3}\,\rm{K}$ (the MW value for the external
pressure). We find that the \CII luminosities predicted when adopting
the various density profiles are all brighter than the observational
constraints. The clear difference between observations and model
predictions increases towards higher redshifts. At $z=0$ the
predictions by the Average, Logotropic, and Powerlaw profile are
relatively similar. The Plummer profile predicts fainter \CII
luminosities. The difference between the various profiles increases
towards higher redshifts. Especially at $z=6$ the model adopting the
Average density profile predicts \CII luminosities that are significantly brighter than the
other three variants. The physical cause of the bright \CII
luminosities is twofold. Firstly, the molecular clouds do not become smaller
and denser in high-pressure environments, resulting in a larger
ionized mass fraction of the cloud. Second, because the clouds are
less dense, the mass of molecular hydrogen within the individual clouds is
lower. The model therefore needs to sample more clouds from the
cloud distribution function in order to equal the molecular hydrogen
mass of the galaxy as calculated in Equation \ref{eq:H2_GK}. This
increases the amount of \CII emission originating from molecular
clouds. At $z=6$ this even leads to unphysical
situations for the model variant adopting the Average density
profile. The total gas mass locked up in molecular clouds that is
necessary to equal the molecular hydrogen mass dictated by Equation
\ref{eq:H2_GK} is larger than the total gas mass of the galaxy as
predicted by the SAM.

For completeness, we present the predicted CO J$=$1--0 through J$=$5--4 luminosity for
$z=0$ galaxies when assuming a constant $P_{\rm ext}/k_{\rm B}
= 10^4\,\rm{cm}^{-3}\,\rm{K}$ in Figure \ref{fig:COz0pressure}. We
find clear differences between the four different molecular cloud
density profiles. The Powerlaw and Plummer density profiles are the
only two that are still in agreement with the observations. The
other two profiles predict CO luminosities that are much
fainter. Especially the Average profile predicts CO luminosities that
are incompatibly low compared to observations. This
difference increases for higher rotational CO transitions, indicating
that the excitation conditions are different  (with a fixed pressure the clouds are less dense and hence the low densities have a stronger effect on the high-J CO lines). We present the CO
luminosity of higher redshift galaxies as a function of their SFR when
assuming a constant $P_{\rm ext}/k_{\rm B} =
10^4\,\rm{cm}^{-3}\,\rm{K}$ in Figure
\ref{fig:COhighzPressure}. Similar to the CO luminosity of $z=0$
galaxies we find that the Powerlaw and Plummer models still reproduce
the observations. When adopting the other profiles the CO luminosities
decrease, especially for the Average density profile. The difference
becomes more dramatic for higher CO rotational transitions. 

For three out of the four (Powerlaw, Logotropic, Average) adopted
density profiles the predicted CO luminosities decreased when fixing
the external pressure to a MW value (most notably for the Logotropic
and Average profile). This is driven by a decrease in the density of
molecular clouds in high-pressure environment, changing the excitation
conditions of CO as well. The Plummer profile variant is the only one for which the CO luminosities
slightly increase when adopting a fixed MW external pressure. The
  reason for this is that the Plummer profile has a long tail
towards larger radii with relatively high densities (a few 1000
$\rm{cm}^{-3}$, see Figure \ref{fig:profiles}). This tail constitutes
a large mass fraction and contributes significantly to the overall CO
abundance within molecular clouds, and hence the CO luminosity
(Figure~\ref{fig:profiles_abund}). As the pressure increases, the fraction of the mass in this tail decreases.

In Figure
\ref{fig:CIpressure}  we present the \CI luminosity of galaxies when assuming $P_{\rm
  ext}/k_{\rm B} = 10^4\,\rm{cm}^{-3}\,\rm{K}$. We find that the \CI luminosities predicted by
the Powerlaw, Logotropic, and Average density profiles are almost
identical. 
We furthermore find that the most actively star-forming galaxies have a \CI 1--0 luminosity slightly brighter than the model variants where the external pressure is not set to the MW value.

Summarizing, we find that the increased external pressure in FIR
bright galaxies leads to fainter predicted \CII and \CI luminosities. It leads to
brighter CO luminosities for the Powerlaw, Logotropic, and Average
density profiles, and fainter CO luminosities for the Plummer
profile. Overall we find that a model assuming a Plummer density
profile where the size of molecular clouds depends on the external
pressure acting on the molecular clouds reproduces best the available constraints for \CII, \CI, and CO at low and high redshifts. In the remaining of the paper we will use the Plummer-Pressure dependent model as our fiducial model to explore other sub-grid variations.

\begin{figure*}
\includegraphics[width = 1.0\hsize]{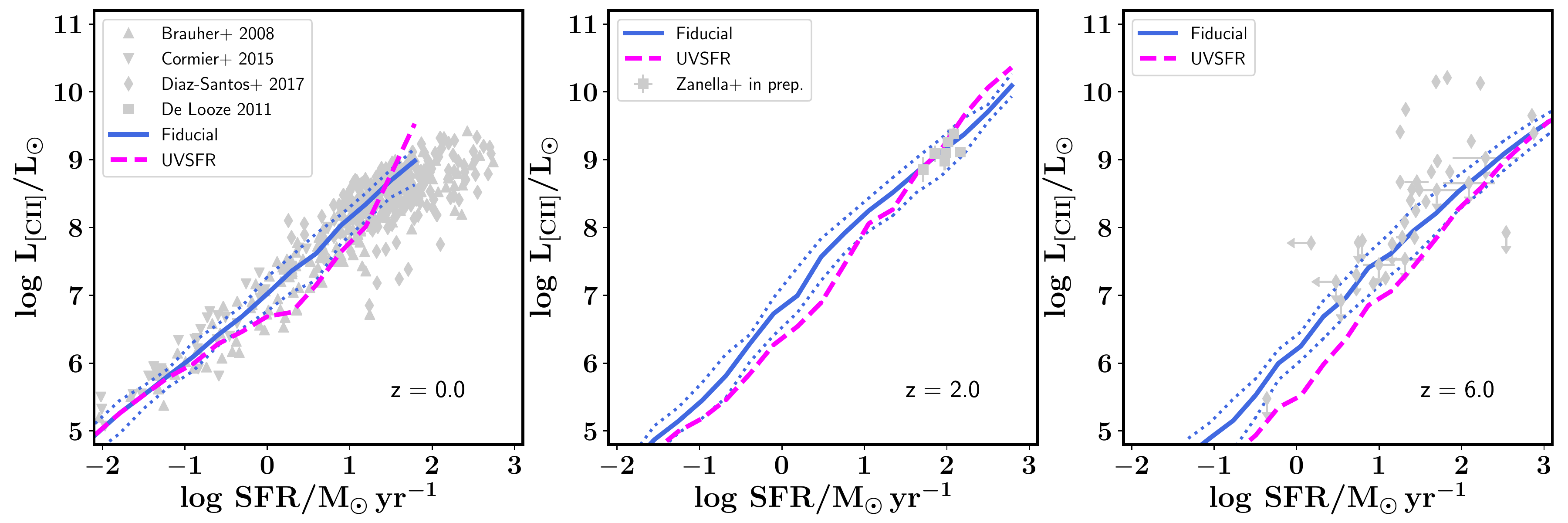}
\caption{The \CII luminosity of galaxies as a function of their SFR at $z=0$, $z=2$, and $z=6$ for a model variant where the
  UV radiation field and CR strength are scaled as a function of the
  local SFR surface density (Fiducial) and as a function of the global
  galaxy SFR (UVSFR). This figure is similar to Figure
  \ref{fig:CIIdensity} aside from the varying relationship
    between the UV field and the SFR imposed in this figure.  It is
    clear that a relationship that ties the UV field to the global SFR
    of galaxies  underpredicts the [CII] luminosity at $z=0$ for galaxies with a
    SFR less than $40\,\rm{M}\,\rm{yr}^{-1}$ and at high-redshift$z=6$.
    It furhtermore predicts \CII luminosities for $z=0$ galaxies with
    SFRs higher than $40\,\rm{M}\,\rm{yr}^{-1}$ that are too bright. Tying
    the UV flux to $\Sigma_{\rm SFR}$ results in predictions in good
    agreement with the observational constraints. \label{fig:CIIradfield}}
\end{figure*}

\subsection{Turbulent compression of gas}
\label{sec:clumping}
Turbulence can cause a non-uniformity of the gas resulting in dense clumps within the ISM. The clumping factor represents the factor by which the mass-weighted mean
density exceeds the volume-weighed mean density and is often
approximated as a function of the Mach number of the gas (the ratio
between the velocity dispersion and sound speed). This has been studied extensively in simulations of turbulent clouds \citep[e.g.,][]{Ostriker2001,Federrath2008}. In \texttt{DESPOTIC} this is accounted for by an enhancement in the rates of all collisional processes \citep[see for details][]{Krumholz2014}. We show the effects
of not including this turbulence dependent clumping factor (i.e.,
clumping factor equals 1) in Appendix \ref{sec:appendix_clumping}. The \CII luminosities of galaxies at $z=0$ 
predicted by the model that does not include turbulent compression of
gas are
$\sim0.3$ dex fainter than the luminosities predicted by our fiducial model variant that does
include turbulent compression of gas. At higher redshifts the
difference is minimal. The CO emission predicted by the model variant
that does not account for turbulent compression of gas are $\sim 0.1$
dex fainter for CO J$=$3--2 and higher rotational transition in
galaxies with SFRs less than 1 $\rm{M}_\odot\,\rm{yr}^{-1}$.

We note that the Plummer profile already guarantees a large range of
densities within a molecular cloud, even without invoking a turbulence
driven clumping factor. For the clumping to make a significant difference, the mass weighted variance in density due to clumping must be larger than the variance implied by the Plummer density profile itself.

\subsection{Molecular cloud mass distribution function}
\label{sec:CDF}
Our model assumes a slope for the molecular cloud mass distribution
function of $\beta = 1.8$. In Appendix
\ref{sec:appendix_cloud_dist_function} we examine the effects of
changing this slope to $\beta = 1.5$ and $\beta = 2.0$, the range
typically found for resolved nearby \citep{Blitz2007,Fukui2008,Gratier2012,Hughes2013, Faesi2018} . We find that the difference between the slope adopted in our fiducial model of $\beta = 1.8$ and $\beta = 1.5$ and $\beta = 2.0$ is negligible \citep{Olsen2017}.

\subsection{UV radiation field and CRs}
\label{sec:UV_radfield}
The UV radiation field and CR field strength acting on molecular
clouds is important for the chemistry. We scale the CR and UV
radiation field with the local SFR surface density. A different
approach seen in the literature scales the CR and UV radiation field
with the integrated SFR of galaxies \citep[normalising the SFR to
  $1\,\rm{M}_\odot\,\rm{yr}^{-1}$, e.g.,][]{Narayanan2017}. Figure
\ref{fig:CIIradfield} shows our predictions for the \CII luminosity of
galaxies for our fiducial model where the UV radiation field is
normalised to the SFR surface density and a model where the UV
radiation field is normalised to the integrated SFR of galaxies. At
$z=0$ and $z=2$ the fiducial model predicts \CII luminosities that are
slightly
fainter in galaxies with a SFR less than $\sim 40
\rm{M}_\odot\,\rm{yr}^{-1}$. The fiducial model predicts fainter \CII
luminosities for more actively star-forming galaxies, due to a quick rise in \CII luminosity as a function of
SFR for the model variant based on the galaxy integrated
SFR. At $z=6$ it becomes clear that a model variant with a UV and CR
field based on the integrated SFR of galaxies predicts a steeper slope
for the \CII -- SFR relation. We find that the model based on the
integrated SFR of galaxies reaches poorer agreement with the $z=0$
observations than our fiducial model, especially for the galaxies with
the brightest FIR luminosities. This said, the prediction for the
\CI and CO luminosities of galaxies between our fiducial model and the
model with CR and UV radiation field based on the integrated SFR are
nearly identical (see Figures \ref{fig:COz0radfield},
\ref{fig:COhighzradfield}, and \ref{fig:CIradfield} in Appendix
\ref{sec:appendix_radfield}).

 To explain why the \CII luminosity varies as a function of
   the UV and CR recipe, whereas the CO and \CI luminosity do not, we focus
   in more detail on the chemistry within molecular clouds. We showed in
   Figure~\ref{fig:radfield_abund}  that as the strength of the
   radiation field increases, a larger fraction of total carbon mass is
   ionized and the fraction of carbon mass that is locked up in CO
   decreases. Based on this alone, one would expect that the \CII
   luminosity arising from a molecular cloud increases, whereas the CO
   and \CI luminosities decrease. The bottom-right panel of Figure
   \ref{fig:radfield_abund} shows that the temperature distribution
   within a molecular cloud changes dramatically as the strength of
   the impinging radiation field increases. A fainter CO or \CI
   luminosity due to lower abundances is (partially) compensated by an
   increase in the temperature and the optical thickness of the cloud. For \CII on the other hand, the combination of a higher gas temperature and a larger \CII abundance
   results in even brighter luminosities. This enhancement in gas temperature is very significant in the regimes where most of the carbon is ionized (i.e., where the \CII abundance is significantly larger than the \CI and CO abundances). We see this in Figure \ref{fig:luminosity_profile}, where we show the cumulative \CII, \CI, and CO J$=$1--0 luminosity profile of a molecular cloud with a changing impinging radiation field (analogue to Figure~\ref{fig:radfield_abund}). Indeed, the \CII emission increases further into the cloud with increasing UV radiation. We find that the total \CI and CO J$=$1--0 luminosity stay constant for $G_0=1$ and $G_0=10$ (and $G_0=100$ for CO J$=$1--0). The final luminosity is reached further within the cloud as the UV radiation increases (due to changes in the abundance).    
 
Our prediction that the CO and \CI luminosity of galaxies stay roughly
the same is in part because of a balance between abundance and gas
temperature, but undoubtedly also by pure chance. A different sub-grid
approach that results in a significantly weaker or stronger UV and CR
radiation field does have the potential to predict CO and \CI
luminosities different from our fiducial model. The reason that a
change in radiation field recipe is more notable in the \CII
luminosities of galaxies is that the increase/decrease in gas
temperature goes hand-in-hand with an increase/decrease of the \CII
abundance.

\begin{figure*}
\includegraphics[width = 1.0\hsize]{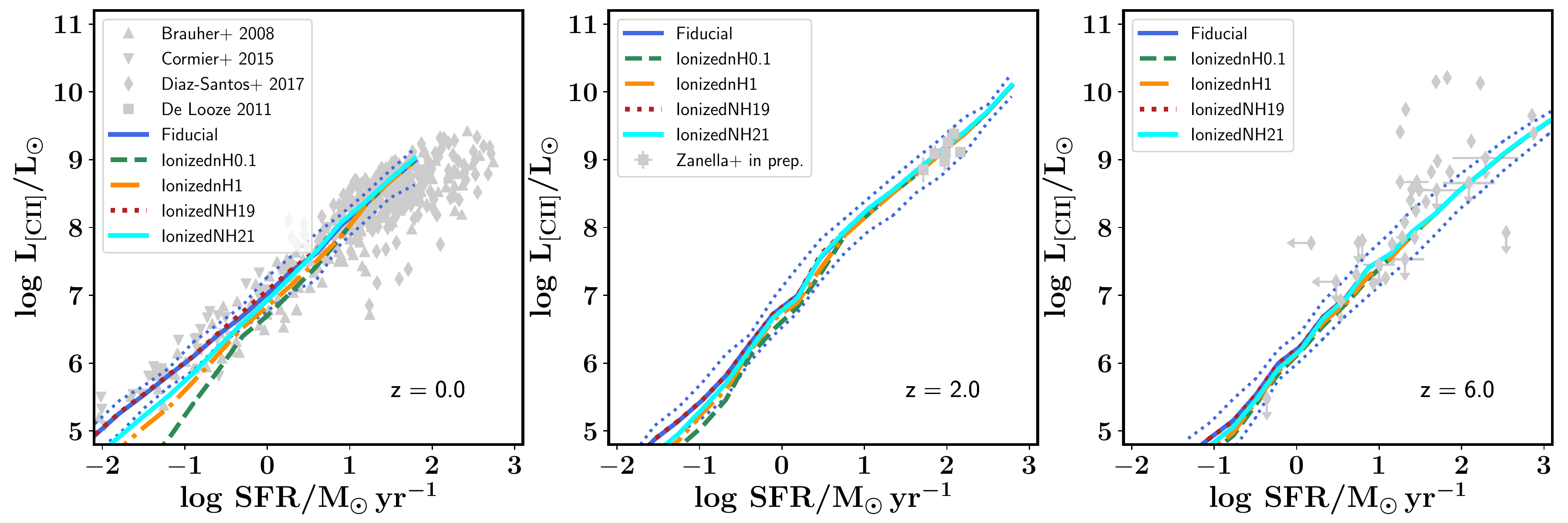}
\caption{The \CII luminosity of galaxies as a function of their SFR at
  $z=0$, $z=2$, and $z=6$ for our fiducial model variant,  variants where the densities of the diffuse ISM are $1\,\rm{cm}^{-3}$ (IonizednH1) and $0.1\,\rm{cm}^{-3}$
  (IonizednH0.1), and variants where the column density are $10^{19}$
  (IonizedNH19) and $10^{21}\,\rm{cm}^{-2}$ (IonizedNH21),
  respectively. This Figure is similar to Figure
  \ref{fig:CIIdensity}. An increase in the density of the atomic
  diffuse ISM results in brighter \CII emission for FIR-faint galaxies at $z=0$, but does not affect the \CII luminosities of $z=2$ and $z=6$ galaxies.\label{fig:CIIionized}}
\end{figure*}

\subsection{Modeling the contribution from diffuse gas}
\label{sec:diffuse}
So far we have focused on the sub-grid choices for the molecular gas in galaxies. The diffuse ISM can also contribute to the \CII emission of galaxies, especially in low-mass and low-SFR galaxies where the diffuse warm ISM constitutes a significant mass fraction of the ISM. Within our fiducial model the atomic gas is modeled as a one-zone cloud with a mass density of $10\,\rm{cm}^{-3}$. In Figure \ref{fig:CIIionized} we show the predicted \CII luminosity of galaxies for our fiducial model and a model variant where we assume the density of the atomic gas to be $1\,\rm{cm}^{-3}$ and $0.1\,\rm{cm}^{-3}$, as well as model variants where we vary the column density of the one-zone clouds from $10^{19}$ to $10^{21} \,\rm{cm}^{-2}$.

We find that lower densities for the atomic hydrogen results in fainter \CII emission for galaxies with low SFRs at $z=0$. We find no significant difference between the different model variants at $z=2$ and $z=6$. This redshift dependence is driven by lower molecular hydrogen fractions in low-mass galaxies at $z=0$ compared to higher redshifts \citep[e.g.,][]{Popping2014,Popping2015}. No differences are found for the \CI and CO emission of galaxies between the different model variants (see Appendix \ref{sec:appendixIonized}). We find identical results between our fiducial model and a variant with a column density of $10^{19} \,\rm{cm}^{-2}$ for the diffuse atomic gas. When adopting a column density of $10^{21} \,\rm{cm}^{-2}$ the model predicts fainter \CII emission in galaxies with SFRs below 1 $\rm{M}_\odot\,\rm{yr}^{-1}$ at $z=0$. At higher redshifts the predicted \CII emission is identical to our fiducial model. As for changing the density of the gas, we find no significant different in the CO and \CI emission of galaxies when adopting a different column density. This indicates that indeed the emission from atomic carbon and CO traces the molecular phase of the ISM. We do acknowledge that our sub-grid model for the atomic and ionized gas is very simplistic, and a more realistic model would account for density variations within the diffuse ISM \citep[e.g.,][]{Vallini2015, Olsen2015CII, Olsen2017}.

\begin{figure*}
\includegraphics[width = 1.0\hsize]{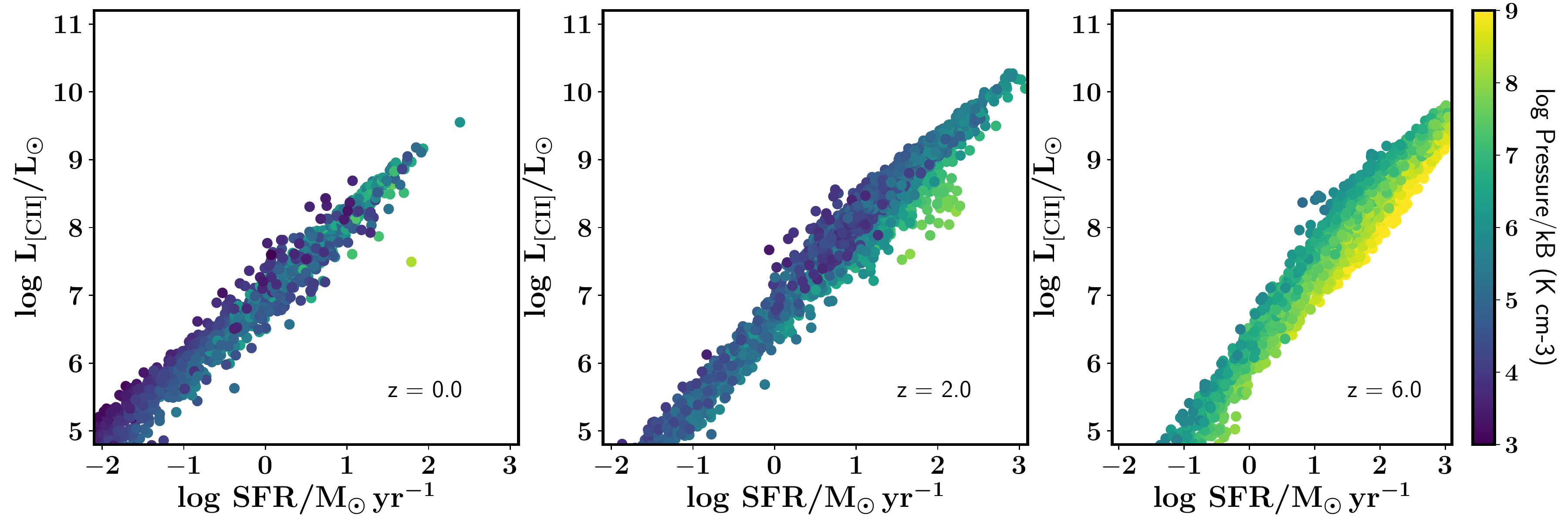}
\caption{The \CII luminosity of galaxies as a function of their SFR at $z=0$, $z=2$, and at $z=6$ for our fiducial model, color coded by the mass-weighted external pressure within galaxies acting on the molecular clouds. Note the clear decrease in \CII luminosity at fixed SFR as a function of increasing pressure. \label{fig:CIIpressurecolor}}
\end{figure*}

\section{Discussion}
\label{sec:discussion}
In this paper we presented a cosmological model that predicts the
\CII, \CI, and CO emission of galaxies. Such models heavily rely on
sometimes uncertain sub-grid choices to describe the ISM. In this work we explored the effects of changing the sub-grid recipes on the \CII, \CI and CO emission of galaxies. We discuss the conclusions that can be drawn from our efforts.

\subsection{Multiple emission lines as constraints for sub-grid
  methods}
Throughout this paper we have compared the predictions by the
different model variants to observations of \CII, \CI 1--0, and
multiple CO rotational transitions. As mentioned before, these
different sub-mm emission lines originate in very different phases of
the ISM, ranging from diffuse ionized gas to the dense cores of
molecular clouds. We have seen that some model variants can for
instance successfully reproduce the \CII emission of galaxies, but
fail to simultaneously reproduce the CO emission of galaxies or the
other way around (where a
model assuming a fixed average density for molecular clouds and no
pressure dependence on the size of molecular clouds most drastically
fails to reproduce the CO luminosities of galaxies). It is only
because multiple constraints are used that we can rule out these
sub-grid model variants. This immediately brings us to the
  critical result of this
paper: \textit{only by using a wide range of sub-mm emission lines
  arising in different phases of the ISM as constraints can the
  degeneracy between different sub-grid approaches be broken.} 

There are additional ways to
  constrain the degeneracy between different sub-grid approaches. Good
  examples of these are spatially resolved observations of individual molecular
  cloud complexes \citep[e.g.,][]{Leroy2017,Faesi2018,Sun2018} and
  high-resolution simulations of molecular cloud
  structures. A clear census of the respective contribution by the
  diffuse and molecular ISM to the \CII emission can be obtained
  through the \NII--to--\CII ratio \citep{Pineda2014,Decarli2014,Cormier2015}. These are
  invaluable additional avenues to constrain the sub-grid methods
  typically adopted for works as presented in this paper.

\subsection{Molecular cloud mass--size relation: the dominant sub-grid component}
In our fiducial model the size of a molecular cloud is set by a
combination of the mass of the molecular cloud and the external
pressure acting on this cloud. A higher external pressure results in a
smaller size and therefore higher overall density within the molecular
cloud. We found that this pressure dependence is essential to
simultaneously reproduce the \CII, \CI, and CO emission of galaxies
over a large redshift range (see Section \ref{sec:pressure}). We
explored this for different radial density profiles for the gas within
molecular clouds and found this statement to be true for all of the
adopted density profiles. Of the four adopted profiles, the model
variant adopting a Plummer density distribution within molecular clouds
is the only one that can simultaneously reproduce the \CII, \CI, and
CO observational constraints. We will use this model variant (Plummer
density profile in combination with a pressure dependence on the size
of molecular clouds) in forthcoming papers to explore the sub-mm line properties of galaxies in more detail.

It is intriguing to realise that the simple recipe we adopted for the
size of molecular clouds in combination with a Plummer density profile
can reproduce the emission of sub-mm lines arising in different phases of the ISM over a large redshift range. We can also phrase this differently: \textit{a key requirement for successfully reproducing the sub-mm line emission of galaxies is a molecular cloud mass-size relation that varies based on the local environment of the molecular cloud} \citep{Field2011,Faesi2018,Sun2018}. 

Besides the importance of the external pressure acting on molecular clouds and the radial density dependence of gas within molecular clouds we have also explored the importance of turbulent gas within molecular clouds, the assumed molecular cloud mass distribution function, and different approaches to model the UV radiation field (and cosmic ray field strength) acting on the molecular clouds. A weaker/stronger radiation field changes the ionization depth within the molecular cloud. In particular we find that a model that scales the impinging radiation field based on the local environment properties (in our case the local SFR surface density) rather than global properties better reproduces the available constraints on the \CII emission of galaxies. We do note that we have not explored `extreme' scenarios where we increase or decrease the CR and UV radiation field strength by orders of magnitude. Such large differences have the potential to also significantly change the atomic carbon and CO abundance of gas within molecular clouds and therefore the resulting \CI and CO emission lines. 

\subsection{Our fiducial model}
In this paper we converged to a fiducial model that best reproduces the \CII, \CI, and CO properties of modeled galaxies within the framework of the underlying semi-analytic model. The key ingredients of this fiducial model include:
\begin{itemize}
\item The density distribution of gas within molecular clouds follows a Plummer profile, such that:
\begin{equation}
n_{\rm H}(R) = \frac{3M_{\rm MC}}{4\pi R_{\rm p}^3}\biggl(1 + \frac{R^2}{R_{\rm p}^2}\biggr)^{-5/2},
\end{equation}
where $R_{\rm p}$ is the Plummer radius, which is set to  $R_{\rm p} = 0.1 R_{\rm MC}$. We account for additional clumping due to turbulence driven compression of the gas (see Sections \ref{sec:DESPOTIC} and \ref{sec:clumping}).
\item The size of a molecular cloud depends on the molecular cloud mass, as well as the external pressure acting on the molecular cloud, such that:
\begin{equation}
\frac{R_{\rm MC}}{\rm{pc}} = \biggl(\frac{P_{\rm ext}/k_{\rm B}}{10^4\,\rm{cm}^{-3}\rm{K}}\biggr)^{-1/4}\biggl(\frac{M_{\rm MC}}{290\,\rm{M}_\odot}\biggr)^{1/2},
\end{equation}
where $k_{\rm B}$ is the Boltzmann constant. 
\item The strength of the impinging UV radiation field scales as a function of the SFR surface density, such that:
\begin{equation}
G_{\rm UV} = G_{\rm UV,MW} \times \frac{\Sigma_{\rm SFR}}{\Sigma_{\rm SFR,MW}}.
\end{equation}
The strength of the CR radiation field also scales as a function of the SFR surface density (see \ref{eq:CR_scaling}).
\item The diffuse atomic gas contributes to the \CII emission of a galaxy and is represented by one-zone clouds with a column density of $N_{\rm H} = 10\times 10^{20}\,\rm{cm}^{-2}$ and a hydrogen density of $n_H = 10\,\rm{cm}^{-3}$.
\end{itemize}

\subsection{Decreasing ratios between \CII and SFR: \CII--FIR deficit}
\label{sec:CIIdeficit}
Observations have suggested that the \CII--FIR ratio of galaxies
decreases with increasing FIR luminosity, such that the FIR-brightest
galaxies ($L_{\rm FIR} > 10^{12}\,\rm{L}_\odot$) have a \CII--FIR
ratio 10 per cent lower than galaxies with fainter FIR luminosities
\citep[commonly known as the \CII--FIR
  deficit;][]{Malhotra1997,Malhotra2001,Luhman1998,Luhman2003,Beirao2010,Gracia2011,DiazSantos2013,
  Croxall2012,Farrah2013}. If we convert FIR luminosity into
a SFR following \citet{Murphy2011}, the same effect can be expected
for the \CII--SFR ratio. An additional interesting feature of the
\CII--SFR ratio, is that many $z\sim6$ galaxies have a
\CII--SFR ratio much lower than one would expect from local \CII--SFR
relations \citep[e.g.,][]{Ota2014,Inoue2016,Knudsen2016}.

We already noted in Section \ref{sec:pressure} that the \CII
luminosity of actively star-forming galaxies is lower for our fiducial model
than a model that assumes a fixed pressure
acting of molecular clouds of $P_{\rm ext}/k_{\rm B} =
10^4\,\rm{cm}^{-3}\,\rm{K}$ (compare Figures \ref{fig:CIIdensity} and
\ref{fig:CIInopressure}). In Figure \ref{fig:CIIpressurecolor} we
show again the \CII luminosity of galaxies as a function of their SFR predicted by our fiducial model. In this case we include a
color coding that marks the mass-weighted external pressure acting on
molecular clouds within each galaxy. We find a clear trend, where at
fixed SFR the \CII--SFR ratio decreases with increasing
external pressure. This is especially clear at $z=2$ and $z=6$, where
the predicted \CII luminosities at fixed SFR can differ as
much as two orders of magnitudes.

A decrease in the \CII--SFR ratio as a function of the external pressure is a natural result of our adopted molecular cloud mass--size relation, which also depends on the pressure acting on the molecular clouds. As the pressure increases, the clouds become smaller and the density increases as well. Because of the higher density a smaller mass fraction of the carbon is ionized, decreasing the \CII luminosity of the galaxies.

This result can (at least partially) explain the observed \CII deficit
of local FIR-bright galaxies \citep[e.g.,][]{DiazSantos2013} and the
large number of non-detection of \CII in $z\sim 6$ galaxies
\citep[e.g.,][]{Inoue2016}. Increased densities in local mergers and
high densities in high-z galaxies (in our framework driven by a high
pressure environment) will naturally result in the \CII deficit and
can explain the non-detections. We will explore this in more detail in
a forthcoming paper, also focusing on variations in the C$^+$
abundance and gas and dust temperatures along the \CII deficit.

\subsection{A comparison to other works in the literature}
\subsubsection{Earlier work by Popping et al.}
\citet{Popping2014} and \citet{Popping2016} also presented predictions
for the CO, \CI, and \CII luminosities of galaxies based on the Santa
Cruz SAM. For clarity we briefly discuss the differences between those
works and the work presented here, both in terms of methodology and
model predictions.

\citet{Popping2014} and \citet{Popping2016} created a
three-dimensional realization of every modeled galaxy, assuming an
exponential distribution of gas in the radial direction, as well as
perpendicular to the galaxy disc. These works employed simple
analytic approaches to calculate the abundance of CO, atomic carbon,
and C$^+$ and the temperature of the gas within every grid-cell of the
three-dimensional realization. These (together with the
density inferred from the exponential distribution) were then used as
input for the radiative transfer calculations. It was assumed that a
grid cell is made up by small molecular clouds all with a size of the
Jeans length that belongs to the typical temperature and density of
the grid cell. Individual molecular clouds were described by a one-zone
cloud with a fixed density, accounting for turbulent compression of
the gas.

The biggest differences in methodology compared to \citet{Popping2014} and
\citet{Popping2016}  are 1) the work presented in this paper only
assumes an exponential distribution in the radial direction and does
not have to make any assumption on the scale length of a galaxy disc
in the z-direction; 2) the molecular mass within a galaxy is made up
by sampling from a molecular cloud mass distribution function; 3)
individual molecular clouds are not treated as one-zone models, but are
allowed to have varying density profiles, 4) we use \texttt{DESPOTIC} to solve for
the carbon chemistry and gas and dust temperatures rather than
adopting simplified analytical solutions. Especially points 2--4 put
the work presented in this paper on a more physics-motivated footing
compared to \citet{Popping2014} and \citet{Popping2016}.

In terms of model predictions the biggest difference is that
\citet{Popping2014} and \citet{Popping2016} were not able to reproduce
the CO, \CI, and \CII emission of galaxies over a wide range of
redshifts simultaneously. Our fiducial model does reproduce these
simultaneously, marking the biggest improvent in model success.

\subsubsection{Other cosmological models for the sub-mm line emission of galaxies}
\citet{Lagos2012} presented predictions for the CO luminosity of galaxies based on a semi-analytic model. The authors parametrize galaxies with a single molecular cloud with a fixed density (a flat radial density profile), UV radiation field, metallicity, and X-ray intensity. They then use a library of radiative-transfer models to assign a CO line-intensity to a modeled galaxy. The biggest difference between their approach and work presented here is that we describe individual galaxies by a wide range of molecular cloud with varying intrinsic properties (density, radiation field, radius). This better captures the different conditions present in the ISM within a galaxy. 

\citet{Lagache2017} used a semi-analytic model as the framework to make predictions for the \CII emission of galaxies. The authors use \texttt{CLOUDY} to calculate the \CII emission of molecular clouds. \citet{Lagache2017} also define a single PDR for each galaxy in their SAM, characterised by a mean hydrogen density (with a flat density profile), gas metallicity, and interstellar radiation field. The authors find \CII luminosities for galaxies at $z>4$ similar to our findings, but have not explored other emission lines and lower redshift ranges.

Besides semi-analytic models a number of authors have made predictions for sub-mm emission lines based on zoom (high spatial resolution) hydrodynamic simulations \citep[e.g.,][]{Narayanan2008, Narayanan2012,Olsen2015CO,Olsen2015CII, Vallini2016,Olsen2017,Pallottini2017,Vallini2018}. \citet{Narayanan2014} also used \texttt{DESPOTIC} to calculate the CO emission from molecular clouds. The authors adopt a flat radial density profile within molecular clouds and adopt a lower-limit in the surface density of molecular clouds. This lower limit automatically ensures a large enough hydrogen/dust column to shield the CO. The authors find that as the SFR surface density of galaxies increases, the CO excitation also changes (higher-J CO lines are more excited). We have not specifically tested this result in our paper, but it is in line with our findings that a high pressure (due to higher gas surface densities which also cause higher SFR surface densities) increases the volume density of the ISM and allows for higher excitation of the high-J CO lines.

\citet{Vallini2018} post-process a zoom-cosmological simulation of one main-sequence galaxy at $z=6$ (spatial resolution of 30 pc) to provide predictions for the CO line emission of this galaxy. Despite the high spatial resolution of this simulation, the authors still need to apply a sub-resolution model for the molecular cloud properties. \citet{Vallini2018} populate a sub-resolution element by individual molecular clouds with densities drawn from a log-normal density distribution with a power-law tail due to self-gravity. The width of the log-normal distribution is set by the Mach number of the gas. 
The CO radiative transfer is then solved using \texttt{CLOUDY}. The authors find that a high gas surface density ($200\,\rm{M}_\odot\,\rm{pc}^{-2}$), combined with a high Mach number and warm kinetic temperature of the gas lead to a peak in the CO SLED at CO J$=$7--6. We have not shown predictions for the CO SLED of galaxies up to this excitation level, but the finding of an increased CO excitation with high gas surface densities and temperatures is in agreement with our general findings. The authors provide very detailed predictions for one object, an approach complementary to the semi-analytic effort focusing on large ensembles of galaxies.

\citet{Vallini2015} presents predictions for the \CII luminosity of $z=6$ galaxies as a function of IR luminosity, in agreement with the observed constraints. The authors find that the \CII luminosity of galaxies at a fixed FIR luminosity decreases as a function of metallicity. On top of this, we argue that changes in the ISM conditions (in our work a denser medium due to an increased external pressure upon molecular clouds) can naturally cause a change in the SFR--\CII ratio of galaxies. 
\citet{Pallottini2017} use the approach developed in \citet{Vallini2015} to make predictions for the \CII emission of a high-resolution zoom-simulation of one galaxy at $z=6$. They find that the \CII luminosity of this single galaxy is in agreement with the upper limits for the \CII luminosity of galaxies based on observations and in the same range as the \CII luminosity predictions of $z=6$ galaxies presented in this work.

\citet{Olsen2015CO} also post-processes a hydro-zoom simulation to calculate the CO emission of three main-sequence galaxies at $z=2$. The authors sample molecular clouds from a cloud-mass distribution function, similar to our approach. The authors then assign a size following a mass-size relation and also adopt a Plummer profile for the radial density distribution of molecular clouds. \citet{Olsen2015CO} finds CO luminosities in agreement with observations and similar to our findings. Like Vallini et al., the Olsen et al. work focuses on the resolved properties of individual galaxies which is a complementary approach to our efforts focusing on large groups of galaxies. \citet{Olsen2015CII} additionally makes predictions for the \CII emission of $z=2$ galaxies based on \texttt{CLOUDY} calculations. The authors predict \CII luminosities similar to our predictions. 

\citet{Olsen2017} presents predictions for the \CII emission of $z=6$ modeled galaxies. Changes with respect to \citet{Olsen2015CII} include  updated \texttt{CLOUDY} calculations and the assumption of a logotropic density profile for the gas within molecular clouds. The authors predict \CII luminosities for $\sim$ 30 galaxies. The predicted \CII luminosity all fall well below expectations based on locally derived relations between star formation rate and \CII luminosity, as well as the predictions by our model. 

\subsection{Caveats}
\subsubsection{The diffuse ISM}
In this work we have implemented a very simplistic model for the
sub-mm line emission arising in the diffuse ISM, consisting of a
one-zone model with a fixed column depth. We demonstrated that
different assumptions for the density of this diffuse gas can affect the \CII emission of galaxies, especially when the ISM is
dominated by this diffuse phase (rather than ISM dominated by
molecular gas, see Section \ref{sec:diffuse}). This immediately
demonstrates the necessity of a more realistic representation of the
diffuse ISM, at least accounting for a range in densities \citep[see
for example][]{Olsen2017}.

\subsubsection{Unresolved galaxies}
One of the intrinsic limitations of the semi-analytic method is the
inability to spatially resolve galaxies. We therefore have to assume a
profile for matter within galaxies, in this paper the commonly adopted
exponential profile. In reality the ISM of galaxies does not
necessarily have to follow an exponential profile, especially in
low-mass galaxies or at very high redshifts. Within our formalism a
more concentrated distribution of gas would immediately increase the
\h2 fraction of the gas within galaxies as well as the pressure acting
on molecular clouds and therefore the density within them. This
naturally changes the carbon chemistry and excitation conditions.

We do want
to emphasize that the choice for an exponential distribution of matter
does not guarantee proper agreement between model predictions and
observations. We furthermore wish to emphasize that a different
distribution of matter within galaxies will also not change the
differences we found between different sub-grid model variants. It is
furthermore important to remember that models that do resolve the
internal structure of galaxies (up to some extent) will have to rely
on the same sub-grid methods as discussed in this work. Furthermore, many of
these models do not reproduce galaxy internal structures \citep[sizes,
surface brightness distribution, see][for a
discussion]{Somerville2014}.

\subsubsection{X-rays and mechanical heating}
 We did not include X-rays as an additional heating source. The heating of X-rays on top of UV radiation and CRs can change the
chemistry and excitation conditions of gas. Studies of the CO spectral
line energy distribution in nearby active galaxies have indeed
revealed strong excitation of high CO rotational transitions \citep[CO
J$=$9--8][]{vanderWerf2010,Meijerink2013}. Since we are only focusing
on CO transitions up to CO J$=$5--4, it is not expected that X-ray heating strongly affects the luminosity of the sub-mm emission lines discussed in this work \citep{Spaans2008}.An additional source of heating not discussed in this work is mechanical heating through shocks \citep{Loenen2008,Meijerink2013,Rosenberg2014a,Rosenberg2014b}.

\section{Conclusions}
\label{sec:conclusions}
In this paper we presented a new cosmological galaxy formation model
that predicts the \CII, \CI, and CO emission of galaxies. We combined
a semi-analytic model of galaxy formation with chemical equilibrium
networks and numerical radiative transfer models. In this paper we
specifically explored how different choices for the sub-grid
components affect the predicted \CII, \CI, and CO emission line
strength of galaxies. Our main conclusions are as follows:

\begin{itemize}
\item It is essential that a wide range of sub-mm emission lines arising in vastly different phases of the ISM are used as model constraints in order to limit the freedom in sub-grid approaches.
\item Small changes in the sub-resolution prescription for the ISM can lead to significant changes in the predicted CO, \CI, and \CII luminosities of galaxies.
\item The key requisite for a model that simultaneously reproduces the
  strength of  multiple emission lines from galaxies in the local and
  high-redshift Universe is a molecular cloud mass--size relation that varies based on the local environment of the molecular clouds (in our framework as a function of the external pressure acting on molecular clouds).
\item A model that scales the impinging UV radiation field and cosmic ray field strength as a function of the local star-formation properties better reproduces the observational constraints for \CII than a model based on the integrated SFR of galaxies. Changes for the \CI and CO luminosity of galaxies are minimal.
\item Not including clumping within molecular clouds and changing the slope for the cloud mass distribution function hardly affect the predicted \CII, \CI, and CO luminosities for our fiducial model setup.
\item A successful model for the \CII emission of galaxies must include a realistic density distribution for the diffuse ISM.
\item A pressure dependence on the size of molecular clouds automatically causes a \CII deficit in high-pressure environments.
\end{itemize}

Our fiducial model successfully reproduces the \CII, \CI, and CO
emission of galaxies as a function of their SFR over
cosmic time  within the context of the current cosmological
model predictions. This fiducial model includes a molecular
cloud mass--size relation that additionally depends on the external
pressure acting on a molecular cloud. It furthermore assumes a Plummer
density profile within molecular clouds, and scales the UV and CR
radiation fields as a function of the local SFR surface density. It
assumes a molecular cloud mass distribution function with a slope of
$\beta = -1.8$ and accounts for turbulence driven clumping within
molecular clouds. Lastly, it assumes a density for the diffuse atomic gas of $10\,\rm{cm}^{-3}$. This fiducial model can be used as a starting point for any group that wishes to model the sub-mm
line emission from molecular clouds in galaxy formation simulations
using a similar approach as presented in this work. Including these kind of approaches in models will increase the constraining power of sub-mm instruments for galaxy formation models and increase the informative role these models can play for future observations. 

The prescriptions presented in this work do not represent a finite list of options. One could think of other approaches with an increasing level of complexity. When exploring other options, one should always take into account that minor changes in the sub-resolution physics can lead to large changes in the resulting emission from galaxies. These can best be constrained when focusing on as many emission lines simultaneously as possible.

\section*{Acknowledgements}
G.P. thanks Karen Olsen for providing a \CII data compilation and for
organizing the 'Walking the line 2018' conference which stimulated the
creation of this work. G.P. thanks Anita Zanella for providing the
\CII luminosities of $z\sim2$ galaxies prior to publication. The
authors thank Romeel Dav\'e, Gordon Stacey, and Chris Faesi for useful discussions and the referee for their comments that improved this work. D.N. was funded in part by grants from the US National Science
Foundation via awards AST-1724864 and AST-1715206 and the Space
Telescope Science Institute via award HST AR-13906.0001 and HST
AR-15043.0001. R.S.S. thanks the Downsbrough family for their generous
support. The simulations in this paper were run on Rusty, supported by
the Center for Computational Astrophysics, Flatiron Institute and on
Draco, supported by the Max Planck Gesellschaft. Part of the writing
of this manuscript was performed during an extended stay at the
Munich Institute for Astronomy and Particle Physics as part of the
workshop 'The Interstellar Medium of High Redshift Galaxies'.




\bibliographystyle{mnras}
\bibliography{references} 

\appendix
\section{The luminosity profile of a molecular cloud}
In this appendix we show the \CII, \CI, and CO J$=$1--0 luminosity profile of a molecular cloud as a function of the impinging radiation field. This is further discussed in Section \ref{sec:UV_radfield}.

\begin{figure*}
\includegraphics[width = 1.0\hsize]{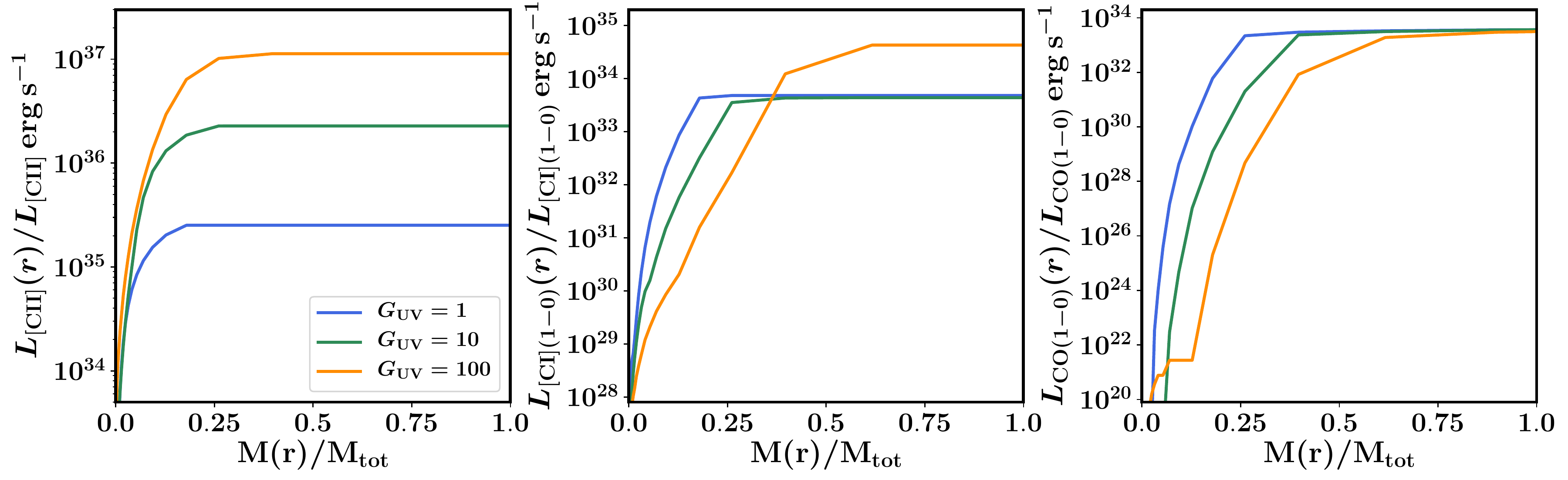}
\caption{The cumulative \CII (left), \CI (center), and CO J$=$1--0 (right) luminosity profile of a molecular cloud for different strengths of the impinging UV radiation field. The molecular cloud has a fixed mass
of $10^5\,\rm{M}_\odot$ distributed following a Plummer density profile, an external pressure acting upon it of $P_{\rm ext}/k_{\rm B} = 10^4\,\rm{cm}^{-3}\,\rm{K}$, and a solar metallicity. As the strength of the UV radiation field increases, the \CII luminosity increases, whereas for instance the total CO J$=$1--0 luminosity stays constant. The final CO J$=$1--0 luminosity is reached further within the cloud as the UV radiation field increases. \label{fig:luminosity_profile}}
\end{figure*}

\section{Clumping}
\label{sec:appendix_clumping}
In this appendix we show the predicted \CII, \CI, and CO luminosities of galaxies for a model variant in which clumping by turbulent gas motions is not taken into account and our fiducial model where this clumping is taken into account.
\begin{figure*}
\includegraphics[width = 1.0\hsize]{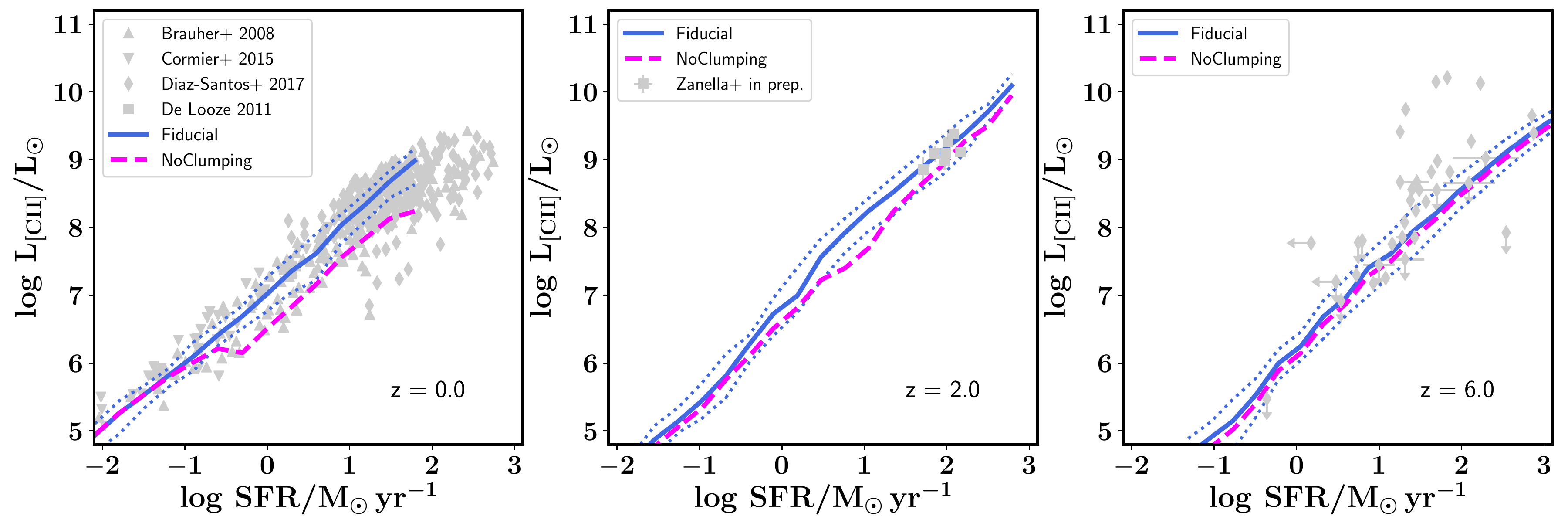}
\caption{The \CII luminosity of galaxies as a function of their SFR at $z=0$, $z=2$, and $z=6$ for a model variant with (Fiducial) and without (NoClumping) turbulence dependent clumping. This Figure is similar to Figure \ref{fig:CIIdensity}. Clumping has a minimal effect on the predicted \CII luminosities of galaxies.\label{fig:CIIclumping}}
\end{figure*}

\begin{figure*}
\includegraphics[width = 1.0\hsize]{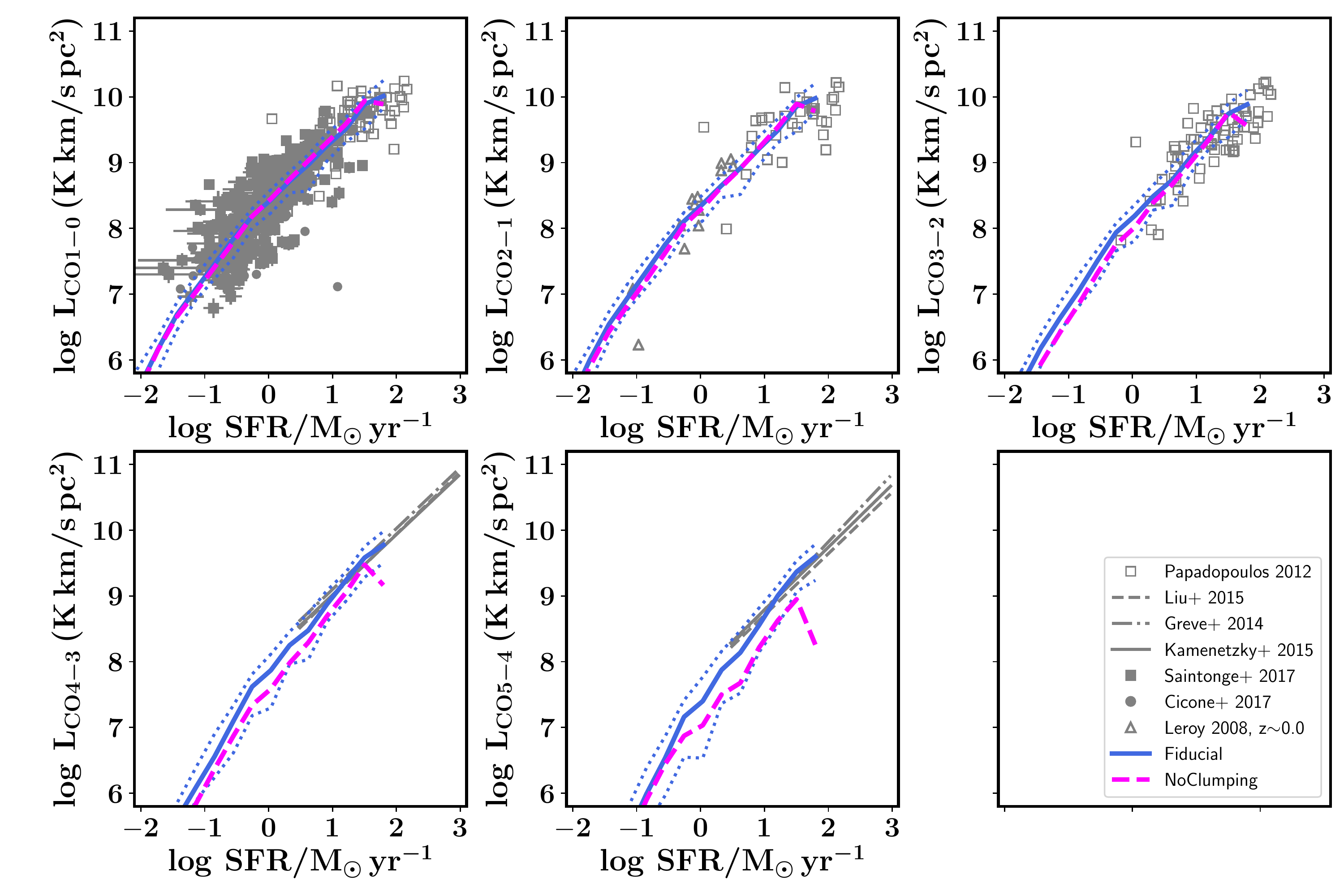}
\caption{The CO J$=$1--0 to CO J$=$5-4 luminosity of galaxies as a function of their SFR at $z=0$ for a model variant with (Fiducial) and without (NoClumping) turbulence dependent clumping. This Figure is similar to Figure \ref{fig:COz0density}. The effect of clumping becomes more important for higher rotational CO transitions, however, this effect is small. \label{fig:COz0clumping}}
\end{figure*}

\begin{figure*}
\includegraphics[width = 1.0\hsize]{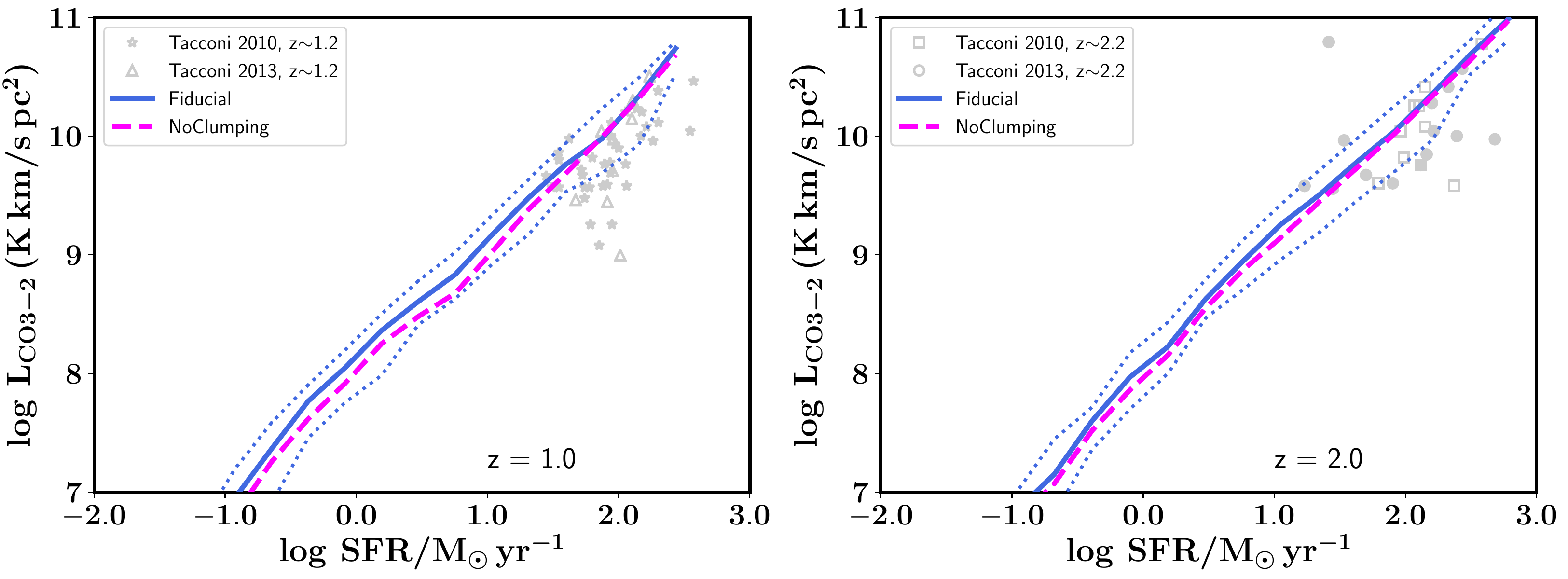}
\caption{The CO J$=$3--2 luminosity of galaxies at $z=1$ and $z=2$ as
  a function of their SFR for a model variant with (Fiducial) and
  without (NoClumping) turbulence dependent clumping. This Figure is
  similar to Figure \ref{fig:COzhighdensity}. \label{fig:COhighzclumping}}
\end{figure*}

\begin{figure}
\includegraphics[width = 1.0\hsize]{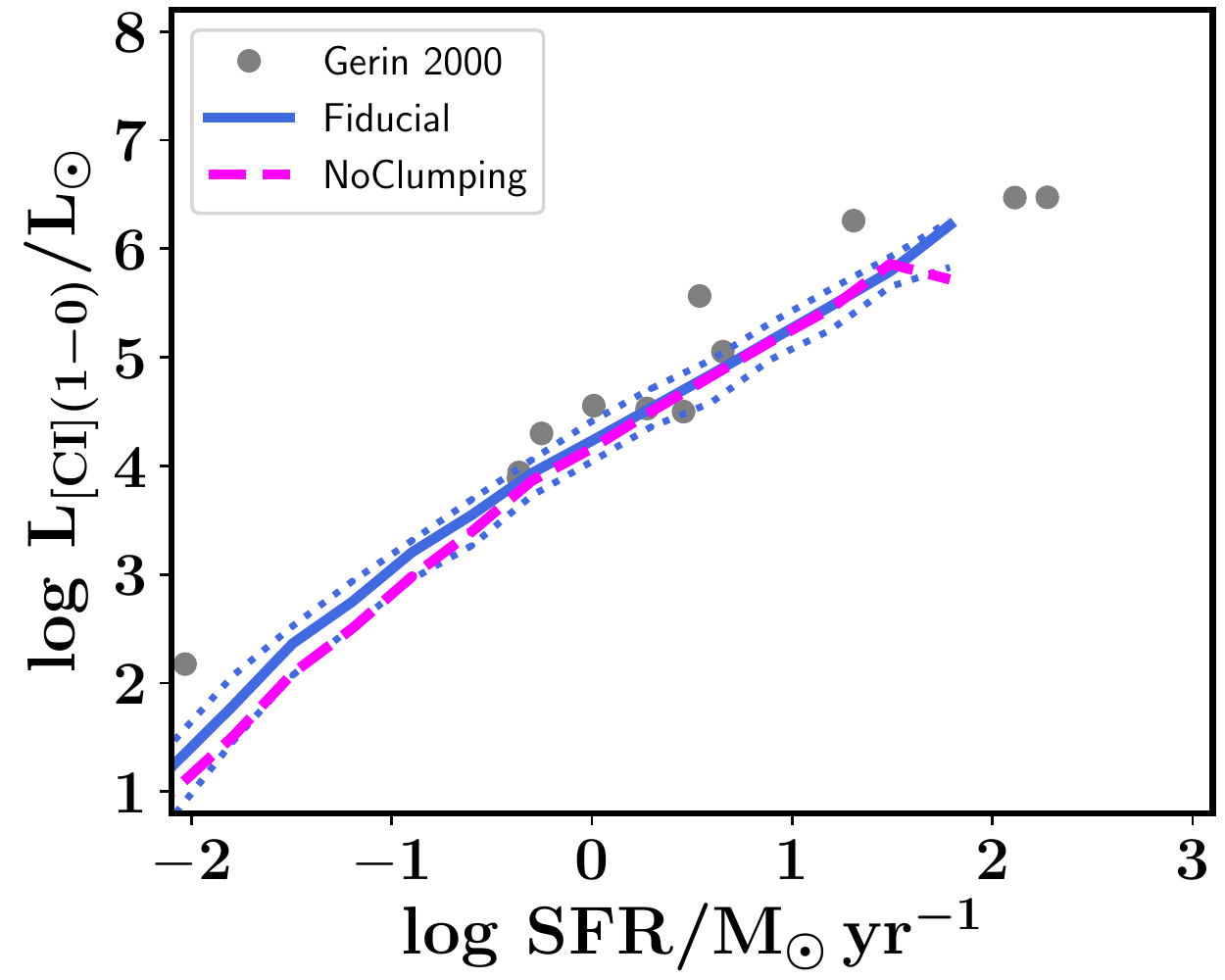}
\caption{The [CI] 1--0 luminosity of galaxies at $z=0$ as a function of their SFR for a model variant with (Fiducial) and without (NoClumping) turbulence dependent clumping. This Figure is similar to Figure \ref{fig:CIdensities}. Clumping has a minimal effect on the predicted \CI luminosities of galaxies.\label{fig:CIclumping}}
\end{figure}

\newpage
\section{Molecular cloud mass distribution function}
In this appendix we show the predicted \CII, \CI, and CO luminosities of galaxies for model variants where we change the slope $\beta$ of the molecular cloud mass distribution function from $\beta$ = -1.5 to $\beta = -2.0$.
\label{sec:appendix_cloud_dist_function}
\begin{figure*}
\includegraphics[width = 1.0\hsize]{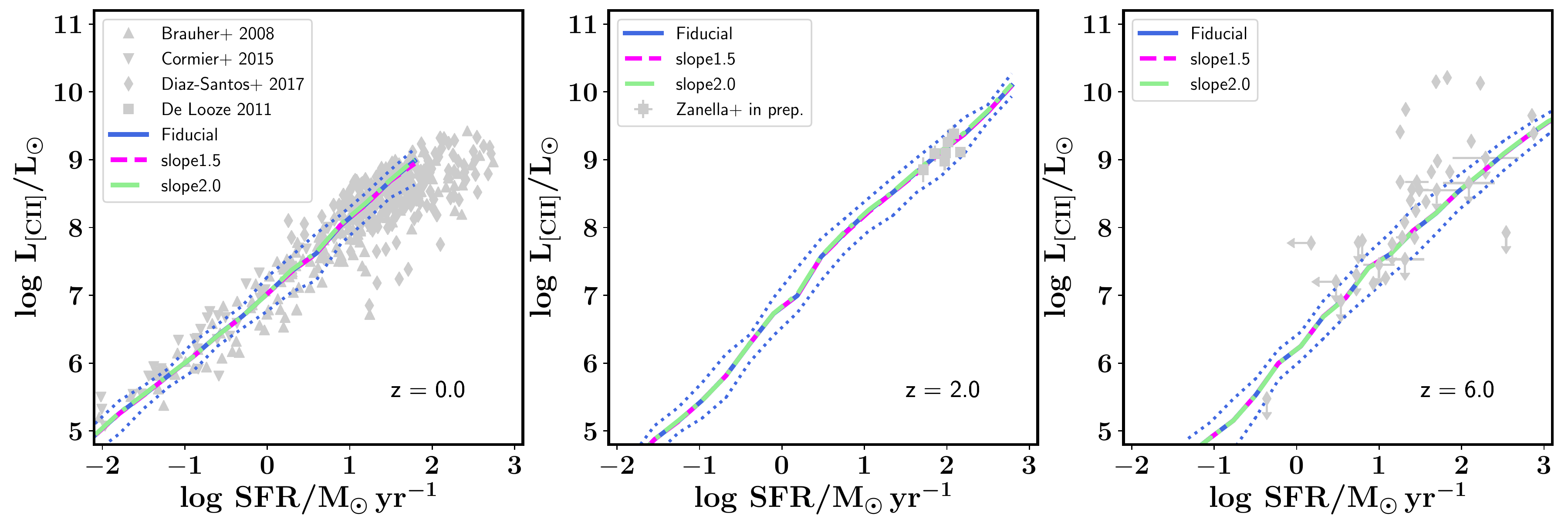}
\caption{The \CII luminosity of galaxies as a function of their SFR at $z=0$, $z=2$, and $z=6$ for a model variant with a slope $\beta$ of the cloud mass distribution function of $\beta=1.8$ (Fiducial), $\beta = 1.5$ (slope1.5)  and $\beta = 2.0$ (slope2.0). This Figure is similar to Figure \ref{fig:CIIdensity}. Different choices for the slope of the molecular cloud mass distribution function do not affect the predicted \CII luminosity of galaxies.\label{fig:CIICDF}}
\end{figure*}

\begin{figure*}
\includegraphics[width = 1.0\hsize]{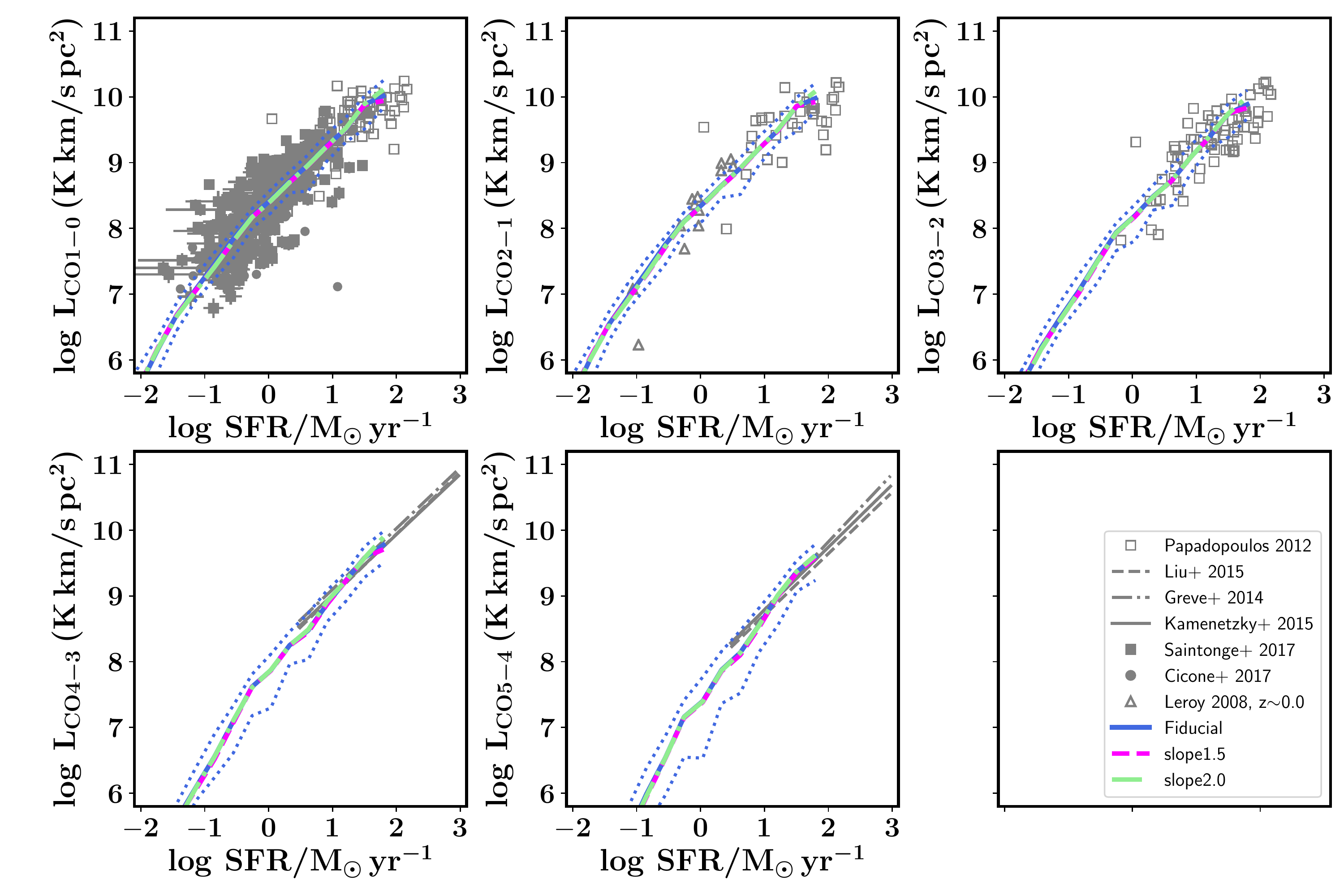}
\caption{The CO J$=$1--0 to CO J$=$5-4 luminosity of galaxies as a function of their SFR at $z=0$ for a model variant with a slope $\beta$ of the cloud mass distribution function of $\beta=1.8$ (Fiducial), $\beta = 1.5$ (slope1.5)  and $\beta = 2.0$ (slope2.0). This Figure is similar to Figure \ref{fig:COz0density}. Different choices for the slope of the molecular cloud mass distribution function do not affect the predicted CO luminosity of galaxies.\label{fig:COz0CDF}}
\end{figure*}

\begin{figure*}
\includegraphics[width = 1.0\hsize]{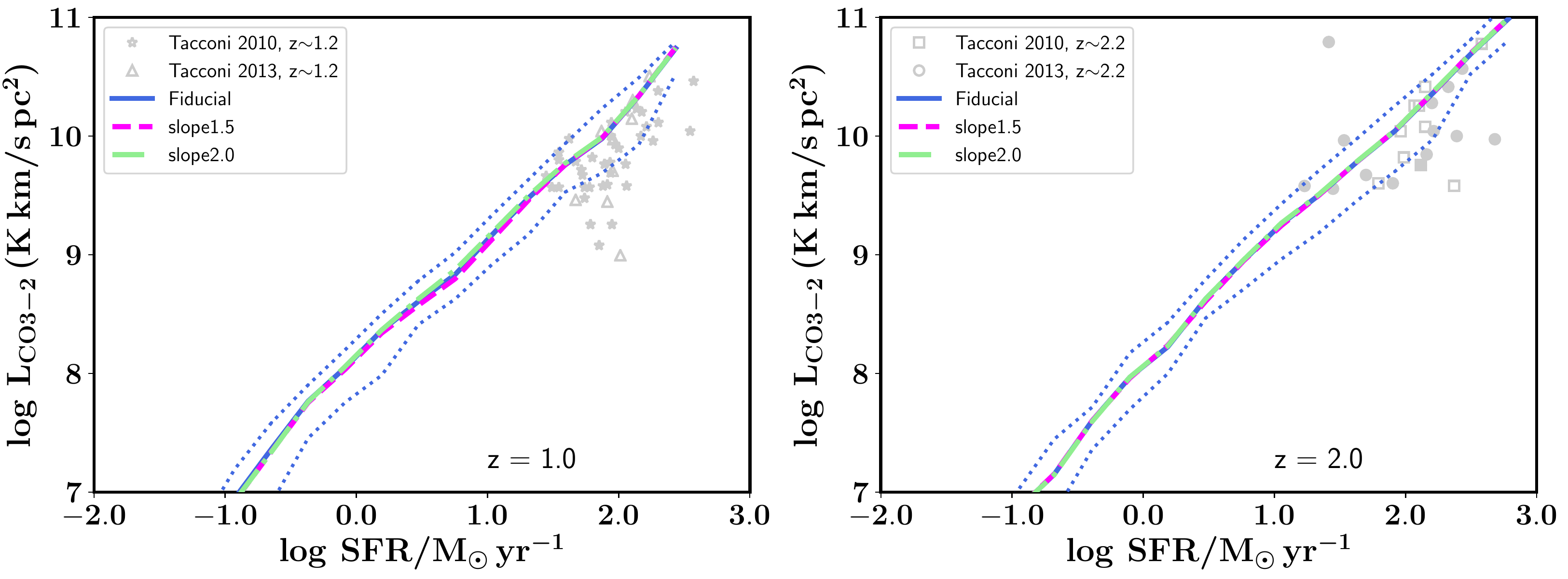}
\caption{The CO J$=$3--2 luminosity of galaxies at $z=1$ and $z=2$ as a function of their SFR for a model variant with a slope $\beta$ of the cloud mass distribution function of $\beta=1.8$ (Fiducial), $\beta = 1.5$ (slope1.5)  and $\beta = 2.0$ (slope2.0). This Figure is similar to Figure \ref{fig:COzhighdensity}. Different choices for the slope of the molecular cloud mass distribution function do not affect the predicted CO luminosity of galaxies.\label{fig:COhighzCDF}}
\end{figure*}

\begin{figure}
\includegraphics[width = 1.0\hsize]{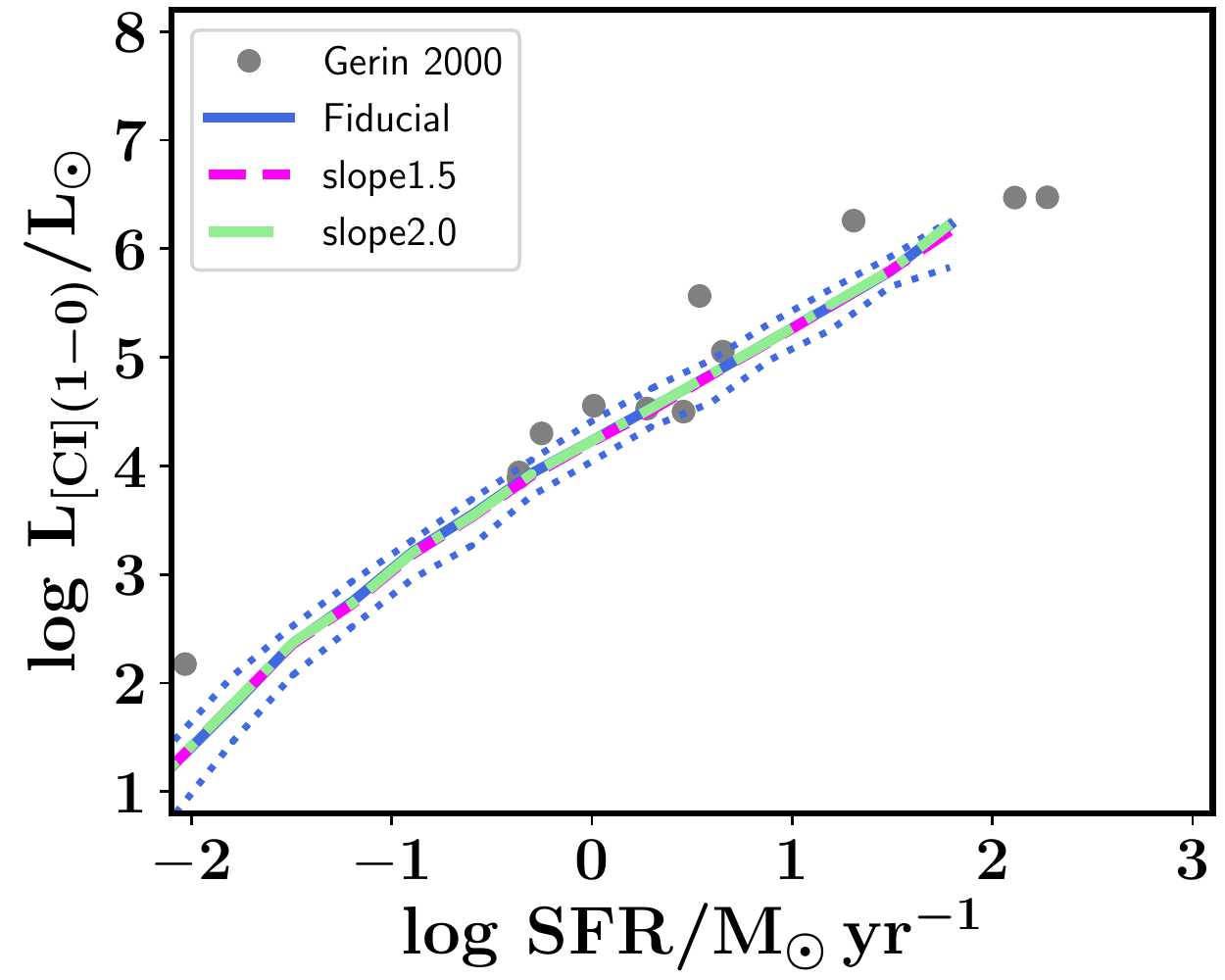}
\caption{The [CI] 1--0 luminosity of galaxies at $z=0$ as a function of their SFR for a model variant with a ope $\beta$ of the cloud mass distribution function of $\beta=1.8$ (Fiducial), $\beta = 1.5$ (slope1.5)  and $\beta = 2.0$ (slope2.0). This Figure is similar to Figure \ref{fig:CIdensities}. Different choices for the slope of the molecular cloud mass distribution function to not affect the predicted \CI luminosity of galaxies.\label{fig:CICDF}}
\end{figure}

\newpage
\section{UV radiation field and CRs}
\label{sec:appendix_radfield}
In this appendix we show the predicted \CI, and CO luminosities of galaxies for our fiducial model where the strength of the UV and CR field scale with the local surface density and a model variant where they scale with the integrated SFR of a galaxy.
\begin{figure*}
\includegraphics[width = 1.0\hsize]{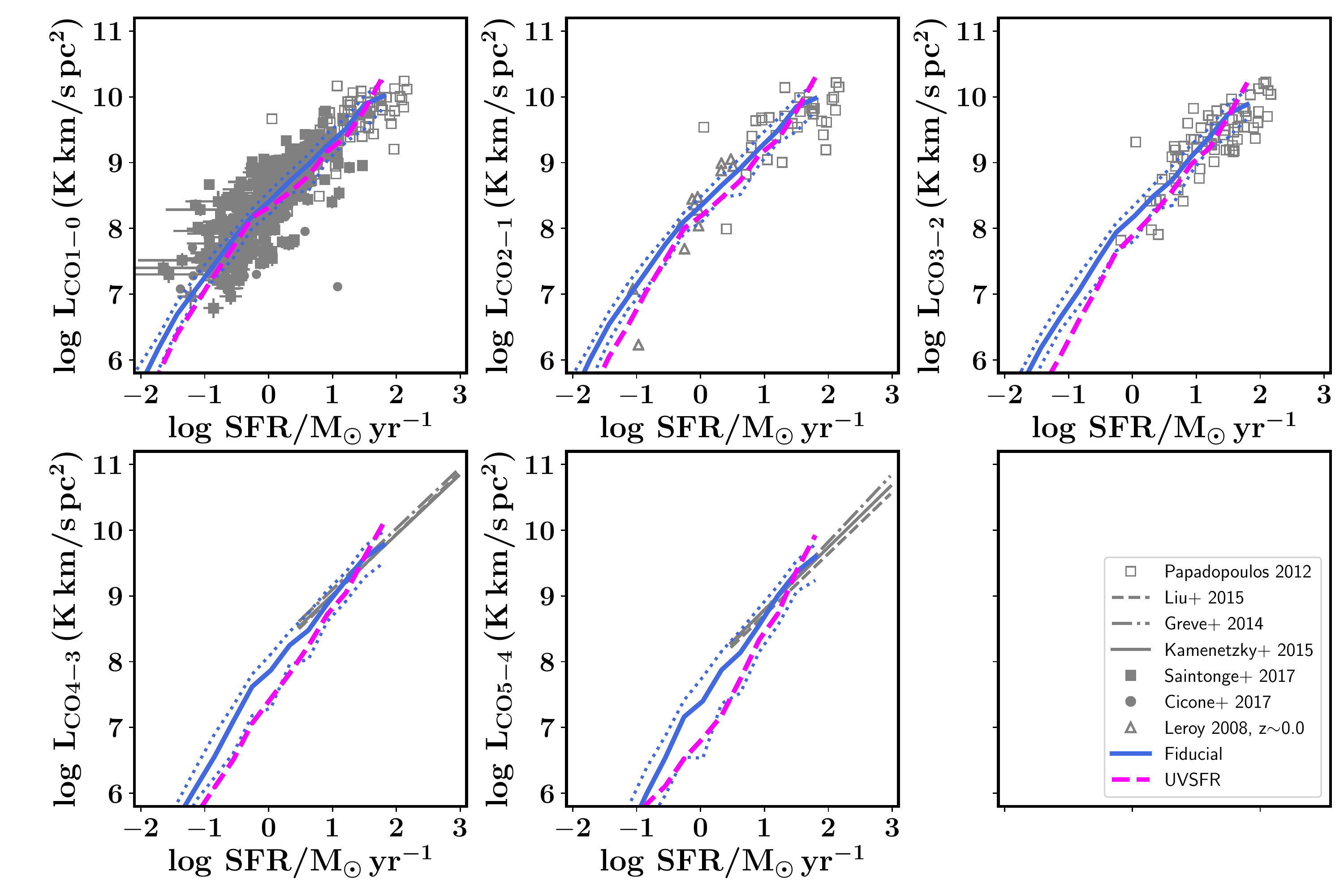}
\caption{The CO J$=$1--0 to CO J$=$5-4 luminosity of galaxies as a function of their SFR at $z=0$ for a model variant where the UV radiation field and CR strength are scaled as a function of the local SFR surface density (Fiducial) and as a function of the global galaxy SFR (UVSFR). This Figure is similar to Figure \ref{fig:COz0density}. The CO luminosities predicted by the two different model variants to scale the UV and CR field are very similar.\label{fig:COz0radfield}}
\end{figure*}

\begin{figure*}
\includegraphics[width = 1.0\hsize]{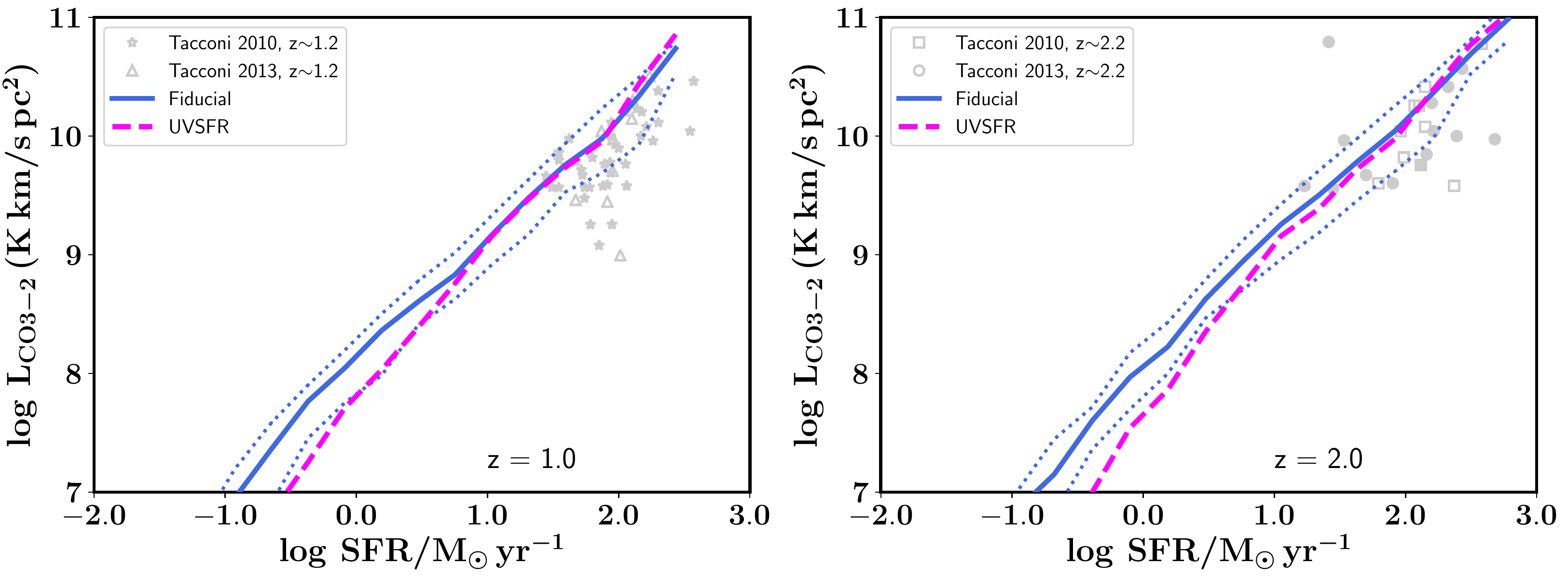}
\caption{The CO J$=$3--2 luminosity of galaxies at $z=1$ and $z=2$ as a function of their SFR for a model variant where the UV radiation field and CR strength are scaled as a function of the local SFR surface density (Fiducial) and as a function of the global galaxy SFR (UVSFR). This Figure is similar to Figure \ref{fig:COzhighdensity}. The CO luminosities predicted by the two different model variants to scale the UV and CR field are very similar.\label{fig:COhighzradfield}}
\end{figure*}

\begin{figure}
\includegraphics[width = 1.0\hsize]{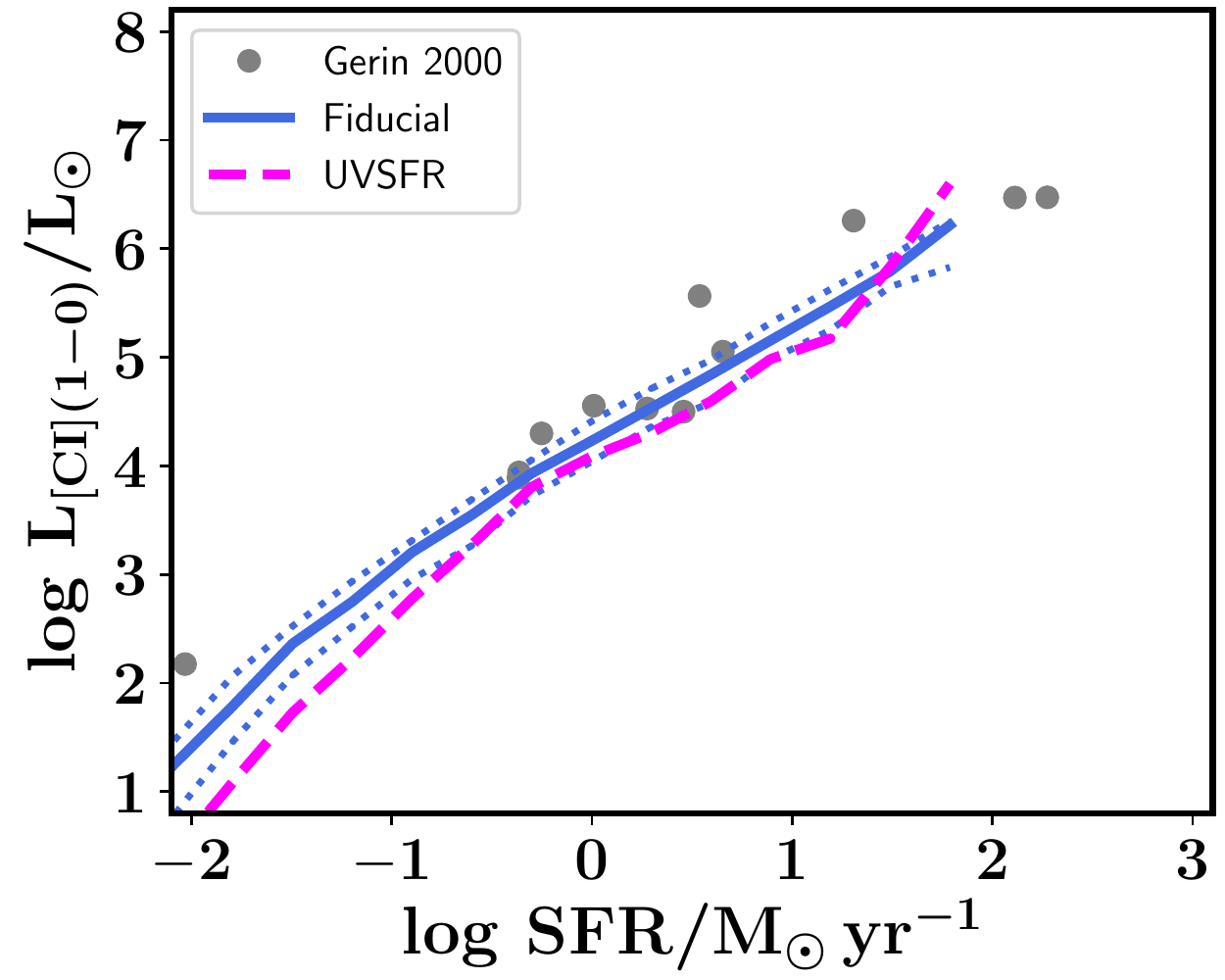}
\caption{The [CI] 1--0 luminosity of galaxies at $z=0$ as a function of their SFR for a model variant where the UV radiation field and CR strength are scaled as a function of the local SFR surface density (Fiducial) and as a function of the global galaxy SFR (UVSFR). This Figure is similar to Figure \ref{fig:CIdensities}. The \CI luminosities predicted by the two different model variants to scale the UV and CR field are very similar.\label{fig:CIradfield}}
\end{figure}

\newpage
\section{Modeling the contribution from diffuse gas}
In this appendix we show the predicted  \CI and CO luminosities of galaxies for model variants where we change the density of the diffuse atomic ISM from 0.1 to $10 \,\rm{cm}^{-3}$.

\label{sec:appendixIonized}
\begin{figure*}
\includegraphics[width = 1.0\hsize]{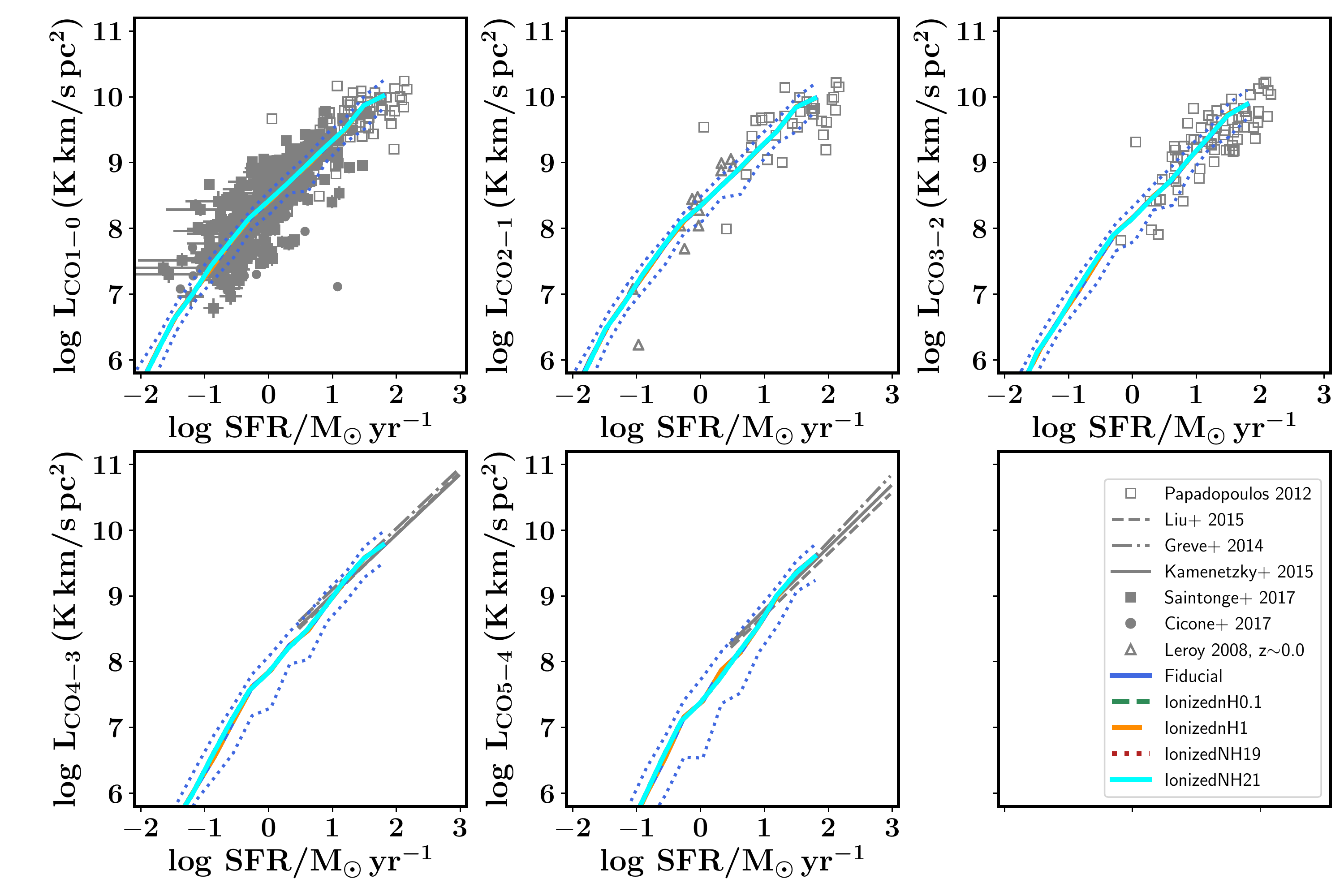}
\caption{The CO J$=$1--0 to CO J$=$5-4 luminosity of galaxies as a function of their SFR at $z=0$  for our fiducial model variant,  variants where the densities of the diffuse ISM are $1\,\rm{cm}^{-3}$ (IonizednH1) and $0.1\,\rm{cm}^{-3}$
  (IonizednH0.1), and variants where the column density are $10^{19}$ (IonizedNH19) and $10^{21}\,\rm{cm}^{-2}$ (IonizedNH21), respectively. This Figure is similar to Figure \ref{fig:COz0density}. The choice for density of the diffuse atomic ISM has no effect on the predicted CO luminosities of galaxies.\label{fig:COz0ionized}}
\end{figure*}

\begin{figure*}
\includegraphics[width = 1.0\hsize]{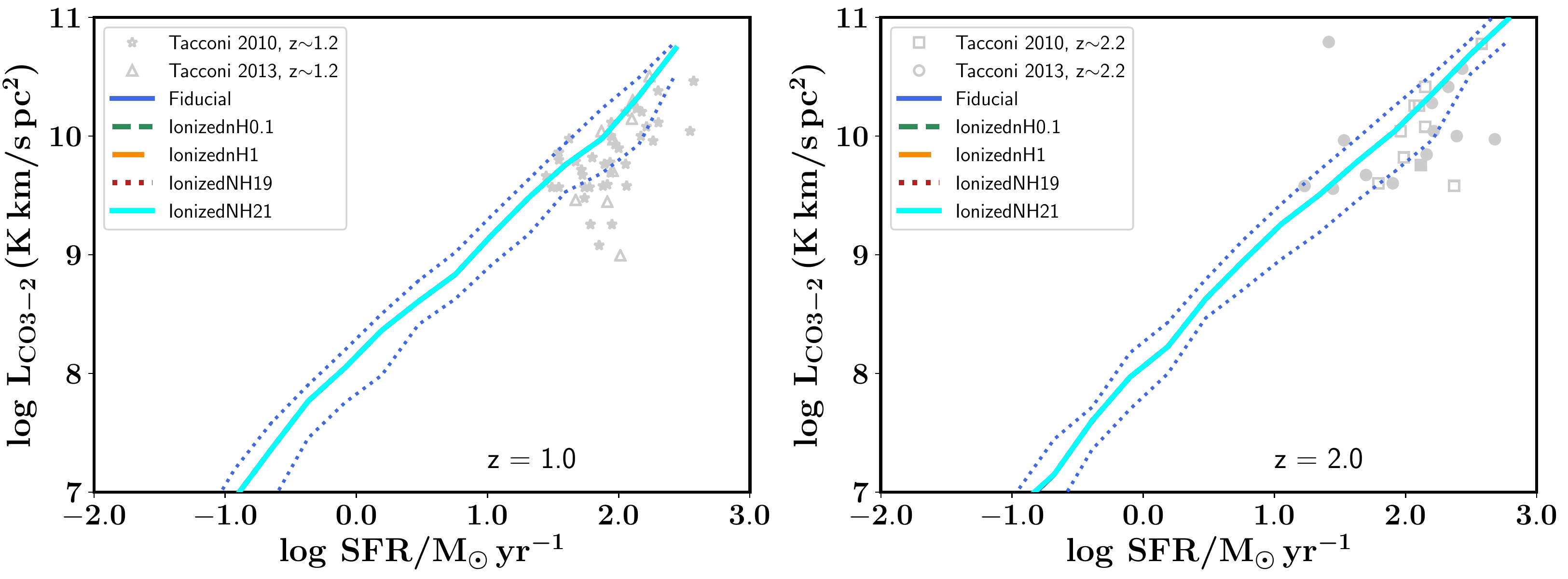}
\caption{The CO J$=$3--2 luminosity of galaxies at $z=1$ and $z=2$ as a function of their SFR  for our fiducial model variant, variants where the densities of the diffuse ISM are $1\,\rm{cm}^{-3}$ (IonizednH1) and $0.1\,\rm{cm}^{-3}$
  (IonizednH0.1), and variants where the column density are $10^{19}$ (IonizedNH19) and $10^{21}\,\rm{cm}^{-2}$ (IonizedNH21), respectively. This Figure is similar to Figure \ref{fig:COzhighdensity}. The choice for density of the diffuse atomic ISM has no effect on the predicted CO luminosities of galaxies.\label{fig:COhighzionized}}
\end{figure*}

\begin{figure}
\includegraphics[width = 1.0\hsize]{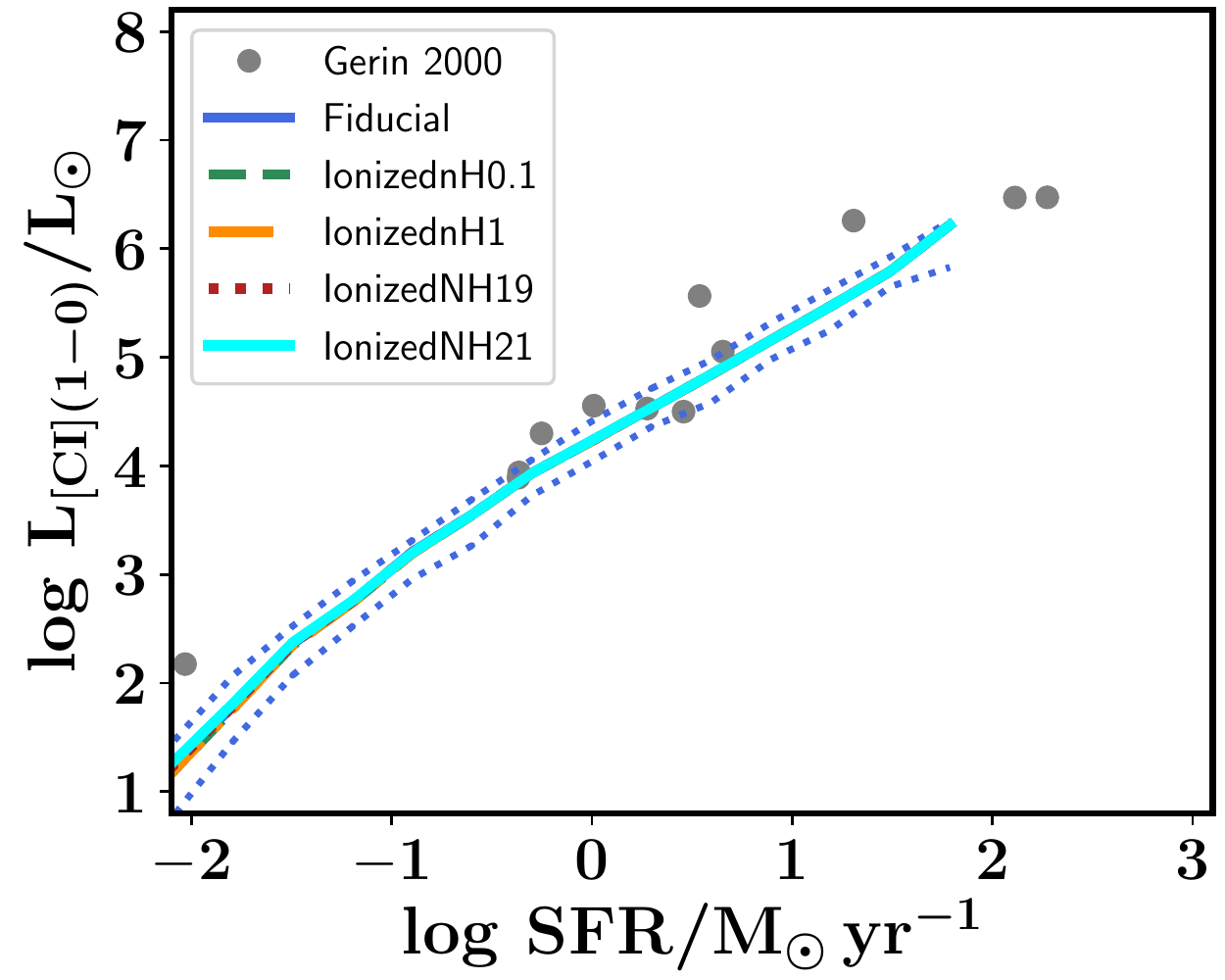}
\caption{The [CI] 1--0 luminosity of galaxies at $z=0$ as a function of their SFR for our fiducial model variant, variants where the densities of the diffuse ISM are $1\,\rm{cm}^{-3}$ (IonizednH1) and $0.1\,\rm{cm}^{-3}$
  (IonizednH0.1), and variants where the column density are $10^{19}$ (IonizedNH19) and $10^{21}\,\rm{cm}^{-2}$ (IonizedNH21), respectively. This Figure is similar to Figure \ref{fig:CIdensities}.The choice for density of the diffuse atomic ISM has no effect on the predicted \CI luminosities of galaxies.\label{fig:CIionized}}
\end{figure}

\bsp	
\label{lastpage}
\end{document}